\documentclass[reprint, amsmath, amssymb, aps, superscriptaddress, prb, longbibliography]{revtex4-2}

\usepackage{graphicx}
\usepackage{epstopdf}
\usepackage{dcolumn}
\usepackage{bm}
\usepackage{color}
\usepackage[colorlinks=true, urlcolor=blue, linkcolor=blue, citecolor=blue]{hyperref}
\usepackage{natbib}
\usepackage{amsbsy}
\usepackage{lipsum}
\usepackage{verbatim}

\usepackage[dvipsnames]{xcolor}
\usepackage{pst-all}
\usepackage{ulem}

\usepackage{lineno}

\begin{document}

\title{
Optical Shubnikov -- de Haas oscillations in 2D electron systems
}	

\author{M.\,L.\,Savchenko}
\affiliation{Institute of Solid State Physics, Vienna University of Technology, 1040 Vienna, Austria}

\author{J.\,Gospodari\v{c}}
\affiliation{Institute of Solid State Physics, Vienna University of Technology, 1040 Vienna, Austria}

\author{A.\,Shuvaev}
\affiliation{Institute of Solid State Physics, Vienna University of Technology, 1040 Vienna, Austria}

\author{I.\,A.\,Dmitriev}
\affiliation{Terahertz Center, University of Regensburg, 93040 Regensburg, Germany}

\author{V.\,Dziom}
\affiliation{Institute of Science and Technology Austria, 3400 Klosterneuburg, Austria}

\author{A.\,A.\,Dobretsova}
\affiliation{Institute of Semiconductor Physics, 630090 Novosibirsk, Russia}

 \author{N.\,N.\,Mikhailov}
\affiliation{Institute of Semiconductor Physics, 630090 Novosibirsk, Russia}

\author{Z.\,D.\,Kvon}
\affiliation{Institute of Semiconductor Physics, 630090 Novosibirsk, Russia}

\author{A.\,Pimenov}
\affiliation{Institute of Solid State Physics, Vienna University of Technology, 1040 Vienna, Austria}
	
\begin{abstract}
\noindent

We report on dynamic Shubnikov -- de Haas (SdH) oscillations that are measured in the optical response, sub -- terahertz transmittance of two-dimensional systems, and reveal two distinct 
types of oscillation nodes: ``universal'' nodes at integer ratios of radiation and cyclotron frequencies and ``tunable'' nodes at positions sensitive to all parameters of the structure.
The nodes in both real and imaginary parts of the measured complex transmittance are analyzed using a dynamic version of the static Lifshitz-Kosevich formula. These results demonstrate that the node structure of the dynamic SdH oscillations provides an all-optical access to quantization- and interaction-induced renormalization effects, in addition to parameters one can obtain from the static SdH oscillations. 

\end{abstract}
	
\date{\today}
\maketitle

Shubnikov -- de Haas (SdH) oscillations are among the most well-known basic phenomena 
reflecting the quantum-mechanical nature of electrons, in particular in two-dimensional electron systems (2DES), where they are a precursor of the quantum Hall effect~\cite{Ando1982}. 
Although the SdH oscillations are thoroughly studied in the static transport response, their observation in optics is limited to several fragmentary measurements~\cite{Abstreiter1976, Fedorych2010, Shuvaev2013, Dziom2019}. 
Because of experimental difficulties, there is no systematic and consistent analysis of such optical Shubnikov -- de Haas oscillations in 2DES so far. 
Whereas, the optical response  represents
a powerful and non-invasive spectroscopic tool to test the disorder and electron-electron correlations in all sorts of two-dimensional materials.

Static and dynamic transport properties are both governed by the frequency-dependent complex conductivity $\sigma(\omega)$. Even in the case of complex optical transmission of the film on a substrate the relation between the measured signal and the conductivity can be written in a simple form, see Eq.~\eqref{eq: TD} below. One may thus expect that the quantum corrections to the conductivity would lead to experimental dependences, which are similar in statics and dynamics. However, already after first treatments of the optical SdH oscillations~\cite{Ando1975,Abstreiter1976} it has been noticed that they have a node near the cyclotron resonance (CR) and reverse their phase around it. Later studies~\cite{Dmitriev2003, Fedorych2010, Raichev2008} have confirmed that the quantum correction to the dynamic conductivity indeed should have an additional modulation governed by the ratio of radiation and cyclotron frequencies.
However, direct evidence for such modulation remained elusive.

Here we report on the observation of the optical SdH oscillations in the transmittance of 2DES.
Two types of nodes can be unraveled there, ``universal'' and ``tunable'' nodes.
We analyse the node structure of oscillations and find that it can be well reproduced using the dynamic version of the Lifshitz-Kosevich formula obtained within the self-consistent Born approximation following Refs.~\cite{Ando1975,Dmitriev2003,Raichev2008}. 

\begin{figure*}
	\includegraphics[width=2\columnwidth]{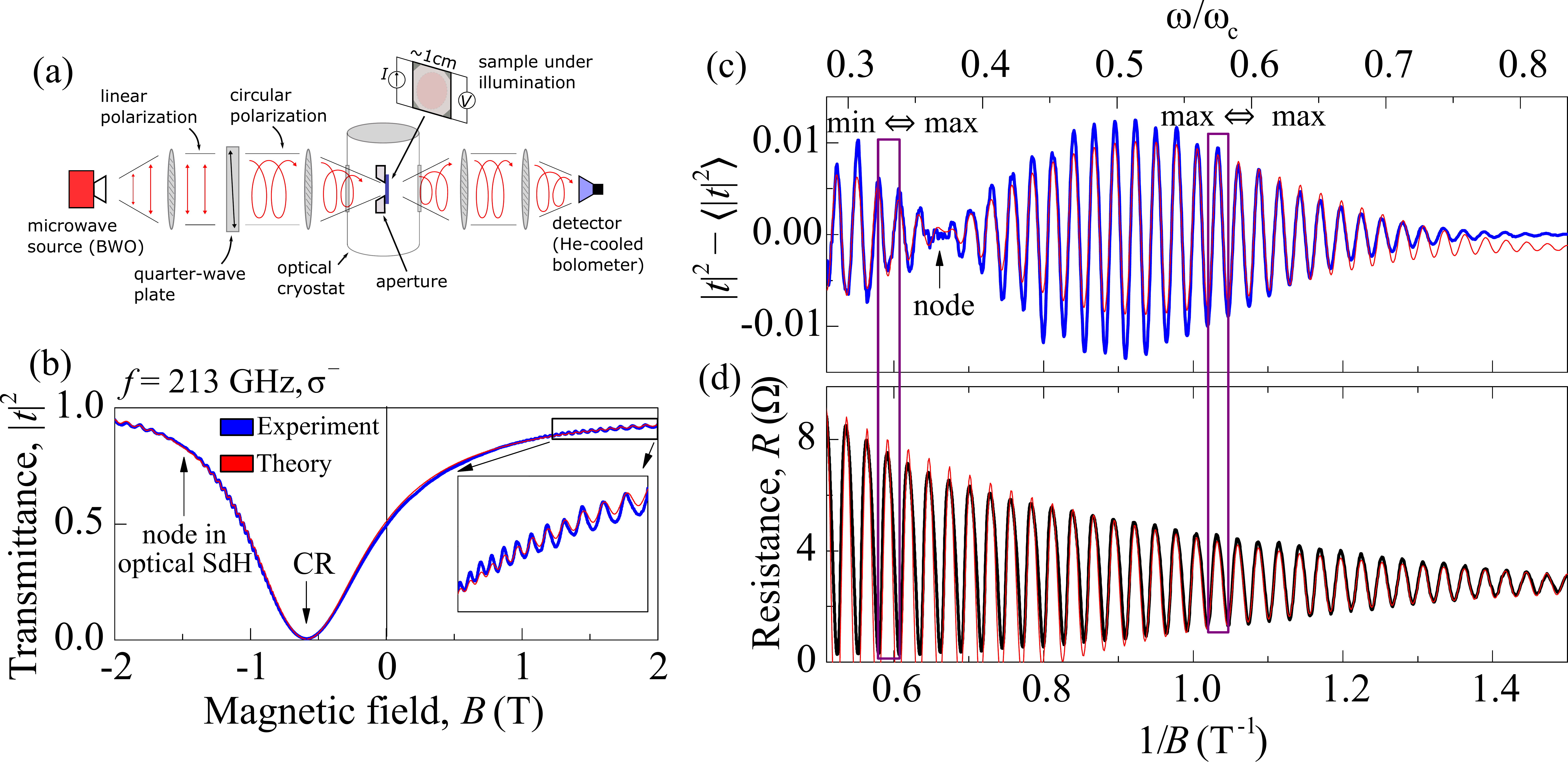}
	\caption{
		(a)~Scheme of the transmission measurements (the irradiated sample area is highlighted).
		(b)~Magnetic field dependence of the transmittance, $|t_-|^2$, measured at $\omega/2\pi=213$\,GHz (blue line).
		(c)~Transmittance oscillations in $1/B$ scale, the smooth part $\langle|t_-|^2\rangle$ of the transmittance is subtracted. Theory curves (red) in~(b) and~(c) are calculated using~Eqs.~\eqref{eq: TD} and \eqref{eq: qc}.
		(d)~SdH oscillations observed in the static longitudinal resistance.
		The node in the dynamic SdH in~(c) separates regions with the same and inverted phase with respect to the transport SdH oscillations shown in (d).
	} \label{fig1}
\end{figure*}

The standard expression for the transmittance of the circularly polarized light through a dielectric slab containing an isotropic 2DES can be written as~\cite{Abstreiter1976, Savchenko2022}
\begin{equation}
\label{eq: TD}
|t_\pm|^2  = \frac{1}{|s_1(1 + \sigma_\pm Z_0) + s_2|^2}.
\end{equation}
Here $\sigma_\pm=\sigma_\text{xx}\pm i\sigma_\text{yx}$ is the dynamic conductivity of 2DES. It is given by the standard Drude expression, $\sigma^\text{D}_\pm = \sigma_0/[1 - i\mu(B_\text{CR} \mp B)]$, in the classical region of perpendicular magnetic field $B$ where the Landau quantization is negligible. Plus and minus signs correspond to the right- and left-handed circular polarization, respectively, $\sigma_0=en\mu$ is the dc conductivity at $B=0$, 
$B_\text{CR}=m_\text{CR} \omega/e$ is the
CR magnetic field defining the CR effective mass
$m_\text{CR}$ of the charge carriers,
$\mu$ is the mobility, 
$n$ is the 2DES density,
and $Z_0 \approx 377\,\Omega$ is the  impedance of vacuum.
Two complex parameters $s_1 = [\cos(kd) - i \epsilon^{-1/2} \sin(kd)]/2$, $s_2 = [\cos(kd) - i \sqrt{\epsilon} \sin(kd)]/2$ describe the Fabry-P\'{e}rot interference in the substrate and are controlled by the product of the sample thickness $d$ and the wave number $k=\sqrt{\epsilon}\,\omega / c$,
where $\epsilon$ is the dielectric constant of the substrate.

Beyond the Drude model, Landau quantization results in the SdH oscillations of the dc resistance, described by the static Lifshitz-Kosevich formula~\cite{Shoenberg1984,Ando1982,Dmitriev2012},
\begin{equation}
\label{eq: lk}
R(B) =R_0 - 4 R_0 \delta \frac{\,\text{X}_T}{\sinh \text{X}_T}\cos\left(\frac{ 2\pi^2\hbar n}{e B}\right),
\end{equation}
where $R_0=R(B=0)$. 
These $1/B$-oscillations are the result of modulation of the density of states, and their period is controlled by the carrier density $n$. At zero temperature, $T=0$, the decay of SdH oscillations at low $B$ is described by the Dingle factor $\delta  = \text{exp}(-\pi / \omega_\text{c} \tau_\text{q})$, where the quantum relaxation time $\tau_\text{q}$ characterises the disorder broadening of Landau levels separated by $\hbar\omega_\text{c}=\hbar e |B|/m$. 
The factor containing $\text{X}_T = 2\pi^2 k_B T/\hbar \omega_\text{c}$ accounts for the additional $T$-smearing. 
In the regime of weak oscillations, where Eq.~\eqref{eq: lk} is valid, transport SdH oscillations provide a powerful and reliable tool to determine such properties of 2DES as the density $n$, single-particle lifetime $\tau_\text{q}$, and effective mass $m$ of charge carriers (entering $\text{X}_T$). In what follows, we present transmission experiments and test the less established, dynamic version of the Lifshitz-Kosevich formula, Eq.~\eqref{eq: qc}, in particular, its nodal structure governed by the ratio $\omega/\omega_\text{c}$.

	\textit{Samples and methods.} 
	2DES with parabolic dispersion was studied in selectively doped 12-nm GaAs quantum well with AlAs/GaAs superlattice barriers grown by molecular beam epitaxy~\cite{Baba1983, Friedland1996, Umansky2009, Manfra2014}.
	The van-der-Pauw sample size was 10$\times$10\,mm, ohmic contacts were placed at the corners.
	After exposure to the room light the electron density and mobility were $n = 1.8\times10^{12}\,$cm$^{-2}$ (only one size-quantized level is occupied) and  $\mu = 2.8\times 10^{5}\,$cm$^{2}$/Vs, respectively.

 2DES with linear dispersion was studied in 6.5-nm HgTe quantum well grown by molecular beam epitaxy~\cite{Kvon2011}.
 The van-der-Pauw sample size was 5$\times$5\,mm.
 A semitransparent 
 Ti/Au 
 gate has been deposited on the 400-nm SiO$_2$/Si$_3$N$_4$ insulator.
 The electron density and mobility at $V_g = 9\,$V were equal to $n = 6.6\times10^{11}\,$cm$^{-2}$ and  $\mu = 5.1\times 10^{4}\,$cm$^{2}$/Vs.
 The Drude optical response of this device was previously studied in Ref.~[\onlinecite{Shuvaev2022}].

		The schemes of our measurements is illustrated in Fig.~\ref{fig1}\,(a) (power transmission, circular polarization) and in Fig.~\ref{fig3}\,(a) (phase-sensitive Mach-Zehnder interferometry, complex transmission amplitude, linear polarization).  
	The samples were irradiated from the substrate side through an 8\,mm (GaAs device) or 4\,mm (HgTe device) apertures.
	Backward-wave oscillators were used as sources of the normally incident continuous monochromatic radiation. 
    The transmittance through the sample was measured using a He-cooled bolometer. 
    The device resistance $R$ (GaAs device) and capacitance $C$ (HgTe device)
	was measured \textit{in situ} using a standard lock-in technique.
	All presented results were obtained at temperature $T=1.9$\,K.

\begin{figure*}
	\includegraphics[width=2\columnwidth]{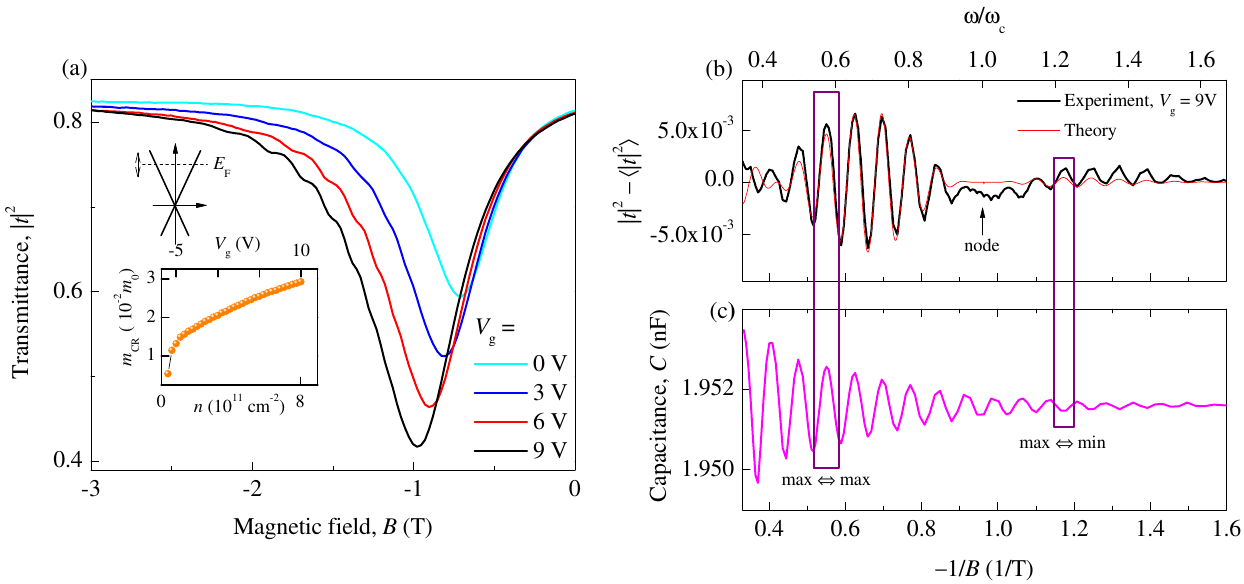}
	\caption{
		(a)~Magnetic field dependence of the transmittance measured at 950\,GHz and with different gate voltages corresponding to different Fermi level positions.
		Dynamic SdH oscillations are more pronounced at higher densities.
		(b)~Transmittance oscillations measured at~(a) 1019\,GHz and at $V_\text{g} = 9\,$V together with SdH oscillations of capacitance~(c), the smooth part of 
		the transmittance $\langle |t|^2 \rangle$ was subtracted. 
		The theory curve (red line) is calculated using Eqs.~(1) and (3) of the main text.
		The phase of dynamic SdH oscillations is flipped across the node.	
		The inset in (a) shows the measured square-root gate voltage dependence of the cyclotron effective mass which corresponds to the linear electron spectrum in the HgTe quantum well with critical thickness. 
	} \label{fig2}
\end{figure*}

\textit{Results and analysis.}
Figure~\ref{fig1}\,(b) shows the magnetic field dependence of the transmittance $|t_-|^2$ measured at $\omega/2\pi=213$\,GHz. Here we studied a GaAs quantum well of high density, $n = 1.8\times10^{12}\,$cm$^{-2}$ (only one size-quantized subband is occupied), and used the left-handed circular polarization, so that only one CR at $B=-B_\text{CR} \approx-0.59\,$T corresponding to the CR mass $m_\text{CR} \approx 0.077 m_0$ is seen, with $m_0$ being the free electron mass. The value of $m_\text{CR}$ is higher than the conduction band effective mass in bulk GaAs, 0.067 $m_0$. This deviation can be attributed to the nonparabolicity of the band dispersion~\cite{Hopkins1987, Kukushkin2015} and to the wave-function penetration into the AlGaAs alloy outside the quantum well; and it is typical for the high-density 2DES in narrow quantum wells.
Apart from the deep CR minimum, the transmittance oscillations are formed at high positive and negative magnetic fields. The oscillations are better seen in~Fig.~\ref{fig1}\,(c), where the smooth part of $|t_- (B)|^2$ is subtracted, and the data are replotted against $1/|B|$. Figure~\ref{fig1}\,(d) shows transport SdH oscillations, measured \textit{in situ}. It is seen that the transmittance and transport oscillations have the same period, which confirms their direct correspondence.

As demonstrated in Figs.~\ref{fig1}\,(c,d), optical and transport SdH oscillations reveal substantial differences as well. The dynamic oscillations have nodes. Here, in~Fig.~\ref{fig1}\,(c), there is one node at $B \approx -1.5\,$T. Across the node, that does not appear in the transport response, the phase of the optical SdH oscillations is flipped, and becomes opposite to that of SdH in $R(B)$ at $|B|$ above the node.
Note that the shape of the SdH oscillations in the static resistance excludes spin splitting \cite{Das1989, Hatke2012} as the origin of the observed nodes in optical SdH oscillations.

In applying the Lifshitz-Kosevich formula, a parabolic dispersion is normally assumed. In order to prove, whether the form of dispersion relations is relevant for the optical SdH oscillations, experiments in a HgTe quantum well of critical thickness $d=6.5$\,nm have been carried out.
This structure hosts Dirac fermions with a linear dispersion and the square root mass-density relation~\cite{Volkov1985,Kvon2011,Buttner2011, Shuvaev2022}. 
In Fig.~\ref{fig2}~(a) the $B$-dependences of the transmittance measured at 950\,GHz and different gate voltages $V_\text{g}$ are shown. 
An increase of the gate voltage results in the increase of the Fermi level position, density, mobility, and the cyclotron mass of the system. This makes the CR minima deeper and wider, and shifts them to higher values of $|B|$. 
The smooth part of these dependences can be well fitted using Eq.~\eqref{eq: TD} and the Drude conductivity. The square root connection between the cyclotron mass and the electron density, see the inset to~Fig.~\ref{fig2}\,(a), confirms the linearity of the spectrum~\cite{Dziom2017, Ikonnikov2011, Shuvaev2022}. 
At high densities the optical SdH oscillations are also seen. 

We compare such oscillations with simultaneously measured static capacitance oscillations in~Figs.~\ref{fig2}\,(b) and~(c). Here, as in GaAs, the period of the optical and static SdH oscillations is the same, and there is a phase flip around the node in the transmittance oscillations.
From positions of the nodes, the CR and effective masses differ by about 4$\%$. This proves the need to include the interaction effects for the Dirac fermions in HgTe quantum wells~\cite{Minkov2020a}.

\begin{figure}		
	\includegraphics[width=1\columnwidth]{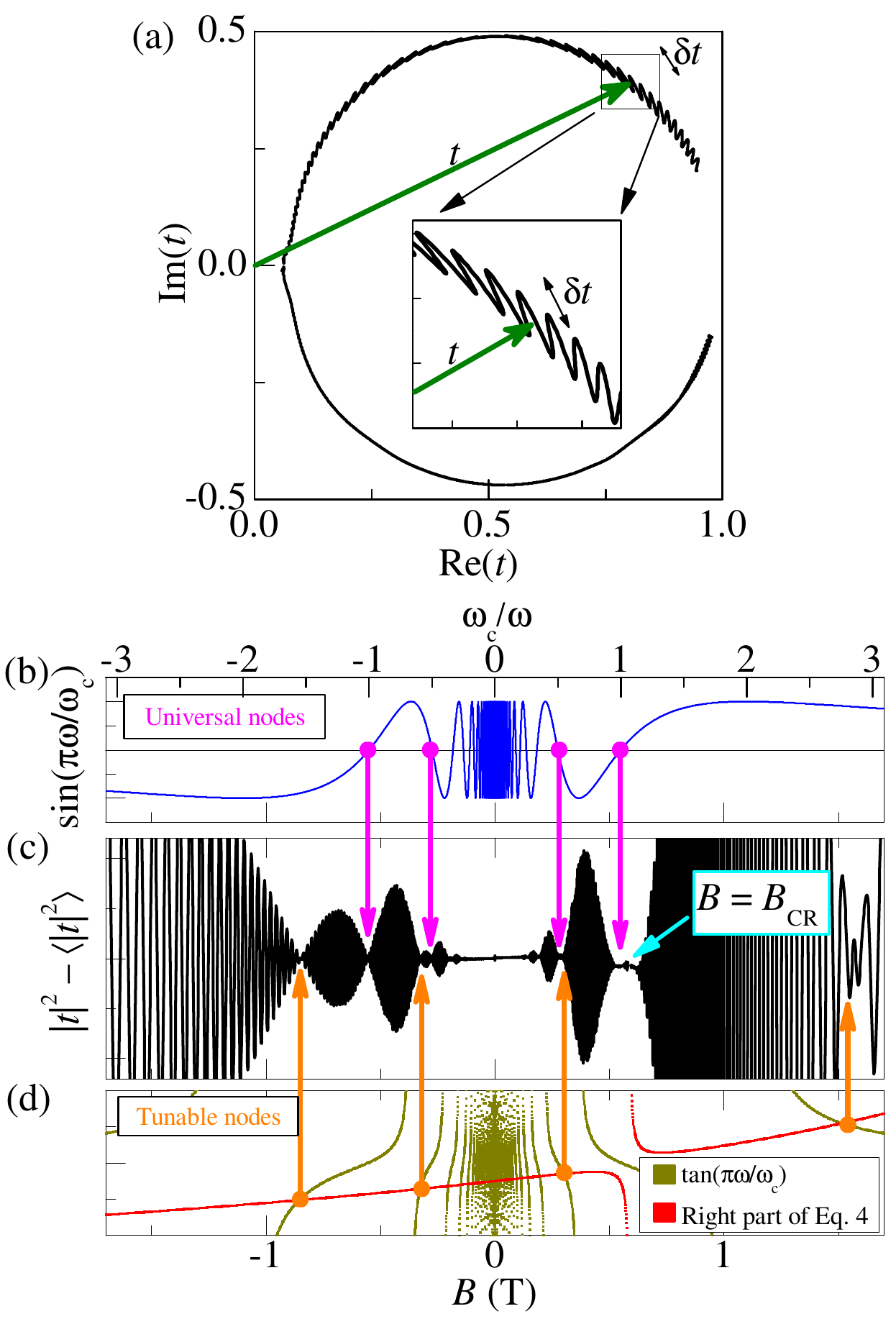}
	\caption{
		Origin of universal and tunable nodes in the transmittance.
		(a)~Imaginary vs real parts of the transmission amplitude $t$. 
		The tunable nodes arise when $t$ and its small correction $\delta t$ are perpendicular to each other on the complex plane.	(b) and (c): Magnetic field dependences of $\sin(\pi \omega / \omega_c)$ (universal nodes) and of calculated transmittance SdH oscillations, see also Supplementary Sec.~\ref{nodes} bellow. (d) Graphical solution of Eq.~\eqref{eq: nodeExpress} determining the tunable nodes in transmittance.
	} \label{Fig_Nodes}
\end{figure}

Our analysis below demonstrates that the observed nodal structure of the optical SdH oscillations can be accurately reproduced using the dynamic version of the Lifshitz-Kosevich formula, Eq.~\eqref{eq: qc}. This formula for the complex dynamic conductivity $\sigma_\pm=\sigma_\text{xx}\pm i\sigma_\text{yx}$, entering Eq.~\eqref{eq: TD}, describes a combined effect of impurity scattering and Landau quantization within the self-consistent Born approximation~\cite{Ando1975}. Previous theoretical treatments of this problem aimed primarily on calculation of the magnetoabsorption, and considered the real dissipative part $\text{Re}(\sigma_\text{xx})$ only~\cite{Ando1975,Dmitriev2003,Fedorych2010}. It has been shown that the dynamic SdH in $\text{Re}(\sigma_\text{xx})$ are modulated as $\sin(2\pi\omega/\omega_\text{c})$, with well-defined nodes at integer and half-integer $\omega/\omega_\text{c}$~\cite{Dmitriev2003,Fedorych2010}. 
As we will see, the nodal structure of the full conductivity $\sigma_\pm$ is more complex.

We analyze the observed optical SdH oscillations using the following dynamic version of the Lifshitz-Kosevich formula \eqref{eq: lk}:
\begin{multline}
\label{eq: qc}
\sigma_\pm = \frac{\sigma_0}{1-i\alpha_\pm}-  4\sigma_0 \delta\frac{ \text{X}_T}{\sinh \text{X}_T}
\cos\left(\frac{ 2\pi^2\hbar n}{e B}\right)\\
\times f(\alpha_\pm) \frac{\omega_\text{c}}{\pi \omega} \sin \left( \frac{\pi \omega}{\omega_\text{c}} \right) \exp \left( \frac{ i \pi\omega}{ \omega_\text{c}}\right).
\end{multline}
Here $f(x) = (1+i/2x)/(i+x)^2$ and $\alpha_\pm=\mu(B_\text{CR}\pm B)$. 
Equation \eqref{eq: qc} is a generalization of the expression for $\text{Re}(\sigma_\text{xx})$ presented in Ref.~\cite{Dmitriev2003}, and can also be extracted from the results of Ref.~\cite{Raichev2008} that considered 2DES with two populated subbands. Similar to Eq.~\eqref{eq: lk}, here it is assumed that the disorder-broadened Landau levels strongly overlap, and only the leading quantum correction, linear in $\delta  = \exp(-\pi / \omega_\text{c} \tau_\text{q}) \ll 1$, is retained. Correspondingly, Eq.~\eqref{eq: qc} is valid away from the CR, $\mu|B_\text{CR}\pm B|\gg \delta$, where such a series expansion is formally justified. However, this does not restrict our analysis below, since we can still rely on the flip of phase of optical SdH oscillations across the node at $\omega=\omega_\text{c}$ in the transmittance data.

In high-mobility 2DES, the parameter $\mu B_\text{CR}$ is usually large. 
Thus, the complex factor $f(\alpha_\pm)$ in Eq.~\eqref{eq: qc} reduces to a real factor $f\simeq\alpha_\pm^{-2}\propto \mu^{-2}$ at all nodes apart from that at $\omega=\omega_\text{c}$.
In this limit, the absorptance given by $\text{Re}(\sigma_\pm)\propto \sin(2\pi\omega/\omega_\text{c})$ is expected to have nodes at both integer and half-integer $\omega/\omega_\text{c}$~\cite{Dmitriev2003,Fedorych2010}. As detailed in Ref.~\cite{Fedorych2010}, the $\omega/\omega_\text{c}$-modulation in this simplest case can be  derived using the Fermi golden rule for the optical transitions, and stems from an oscillating product $\nu(\varepsilon) \nu(\varepsilon + \hbar \omega)$ of initial and final density of states for transitions between disorder-broadened Landau levels.

In contrast to the absorptance, the optical SdH oscillations in transmittance are determined by quantum correction to full complex conductivity which does not vanish at half-integer $\omega/\omega_\text{c}$. The reason is that, unlike the real part, quantum corrections to the full $\sigma_\pm$ cannot be expressed solely through the oscillating density of states: They also explicitly include oscillatory energy renormalization terms originating from the interplay of Landau quantization and disorder~\cite{Dmitriev2003,Raichev2008}.

\begin{figure*}		
	\includegraphics[width=2\columnwidth]{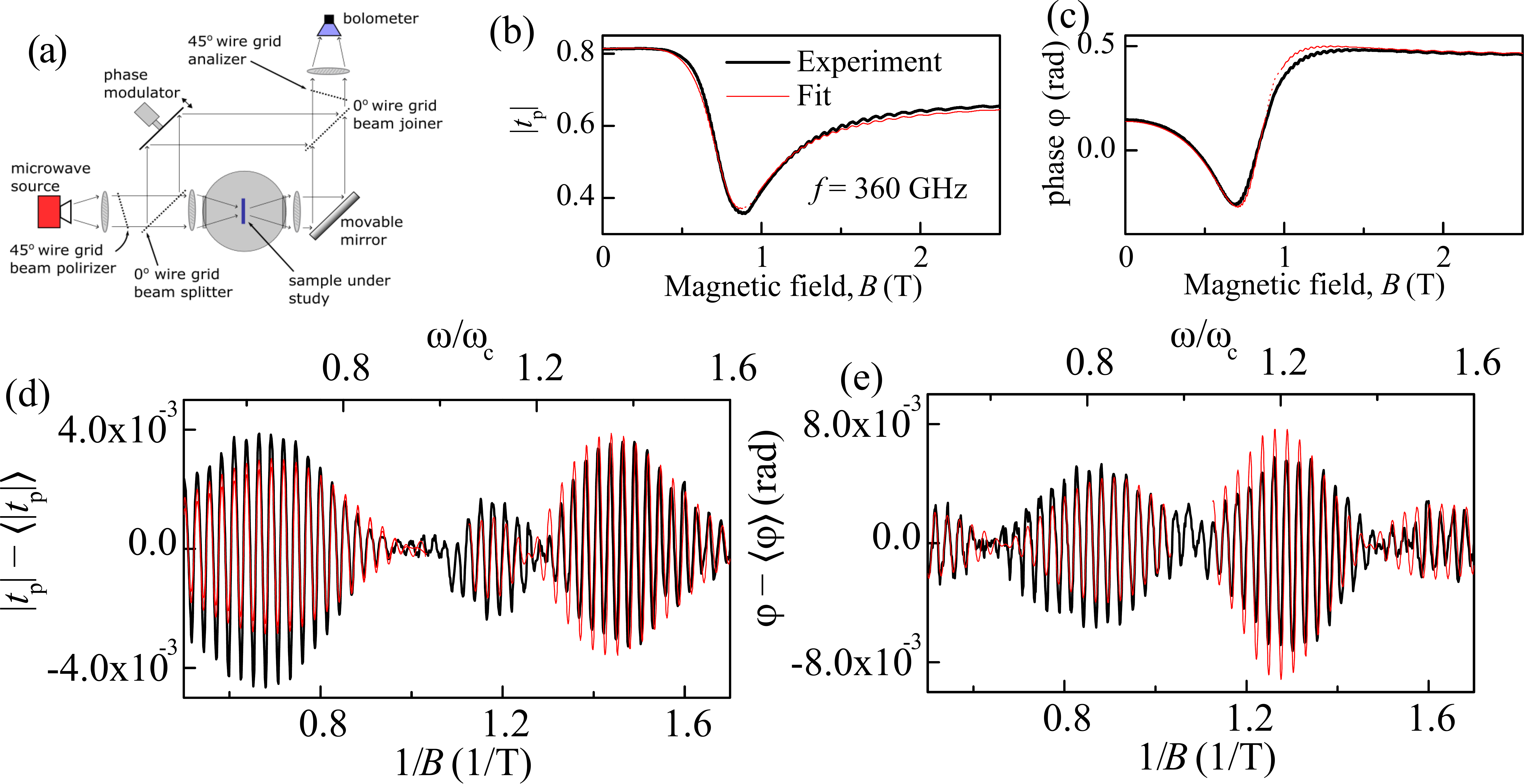}
	\caption{
		(a)~Mach-Zehnder interferometer arrangement for the phase measurements.
		(b)~and (c) Magnetic field dependences of the amplitude of the parallel transmittance $|t_\text{p}|$ and its phase $\varphi$, respectively, measured at 360\,GHz.
		(d)~and (e)~Oscillations of the parallel transmittance amplitude and its phase in $1/B$ scale, the smooth parts in $\langle |t_\text{p}|^2 \rangle$ and $\langle \varphi \rangle$ were subtracted. 
		Black thick lines are experimental data, red thin lines are fits based on~Eq.~\eqref{eq: TD} and \ref{eq: qc}.
	} \label{fig3}
\end{figure*}

As a result, the SdH oscillations in transmittance possess two kinds of nodes, illustrated in Fig.~\ref{Fig_Nodes}. 
First, these are nodes at integer $\omega/\omega_c$, where the oscillatory part of the conductivity Eq.\,\eqref{eq: qc} vanishes. These nodes, which we call universal, appear in all optical measurements: transmission, reflection, and absorption. By contrast, the half-integer absorption nodes do not arise in the transmittance signal. 
At the same time, additional nodes emerge due to the fact that both the Drude transmittance amplitude $t_\text{D}= |t_\text{D}|e^{i\varphi_\text{D}}$ and its quantum correction $\delta t$ are complex numbers. Thus they can be perpendicular to each other on the complex plane, see Fig.~\ref{Fig_Nodes}~(a). Taking into account that $|\delta t| \ll |t_\text{D}|$ while $f(\alpha_\pm)$ is approximately real away from $B=B_\text{CR}$, from Eqs.~\eqref{eq: TD} and \eqref{eq: qc} one obtains that additional nodes appear when (see Supplementary Sec.~\ref{nodes} bellow)
\begin{equation}
\label{eq: nodeExpress}
\tan \left(  \frac{\pi\omega}{\omega_c} \right) = \frac{\sqrt{\epsilon} + \tan(\varphi_\text{D})\tan(kd)}{\sqrt{\epsilon}\tan(\varphi_\text{D}) - \tan(kd)}.
\end{equation}
It is immediately seen that the positions of these nodes can be optically tuned by changing the Fabry-P\'{e}rot  phase $kd$, therefore we call them tunable nodes. 
Graphical solution of Eq.~\eqref{eq: nodeExpress} is illustrated in Fig.~\ref{Fig_Nodes}~(d).

Fig.~\ref{fig1} demonstrates that the theory curves (red lines), calculated using Eq.~\eqref{eq: TD} and \eqref{eq: qc}, closely reproduce our experimental observations, including the formation of the nodes in the transmittance SdH oscillations. The position of the nodes determines the value of the quasiparticle effective mass $m$ that enters Eq.~\eqref{eq: qc} through $\omega_\text{c} = e |B|/m$. All other parameters entering Eqs.~\eqref{eq: TD} and \eqref{eq: qc}, including $\tau_q$, can be obtained from the static resistance, see Fig.~\ref{fig1}~(d), and from the shape of smooth classical transmittance on top of which small quantum oscillations are formed.
Small deviations between theory and experiment can be partially attributed to transition to the separated Landau levels regime, where higher expansion terms in $\delta$ should also be included into the theory. Such deviations are also seen in Fig.~\ref{fig1}~(d) at $B \gtrsim 1$~T where the amplitude of the static SdH oscillations starts to deviate from the cosine-like Eq.~\eqref{eq: lk}.

From the positions of the nodes, for GaAs sample in Fig.~\ref{fig1} we obtain $m = 0.073\,m_0$, about 5$\%$ lower than the CR mass $m_\text{CR} \approx 0.077 m_0$, obtained from the position of the CR minimum. In line with the previous studies~\cite{Hopkins1987, Kukushkin2015, Hatke2013, Tabrea2020, Savchenko2020b}, we attribute this difference to the effective mass renormalization due to electron-electron interactions. 
Such optical experiments provide 
an access to quantization- and interaction-induced 
renormalization effects in 2DES \cite{Tabrea2020, Savchenko2020b}.

We further test the validity of the dynamic  Lifshitz-Kosevich formula, Eq.\,(\ref{eq: qc}), for the phase measurements. Our Mach-Zehnder interferometer setup provides an opportunity to simultaneously measure real and imaginary parts of the complex transmittance amplitude~\cite{Volkov1985a, Shuvaev2012, Dziom2017a, Dziom2018}.
The measurements were performed in configuration with two parallel wire grid polarizers before and after the sample, "beam splitter" and "beam joiner" in~Fig.~\ref{fig3}\,(a).
In this way, we obtained both the absolute value $|t_\text{p}|$ and phase $\varphi$ of the parallel transmittance amplitude $t_\text{p} = |t_\text{p}| e^{i \varphi}$ describing the part of transmitted radiation field with the same linear polarization as in the initial beam. In terms of circular transmission amplitudes entering Eq.~\eqref{eq: TD}, $t_\text{p} = (t_+ + t_-)/2$. 
The interferometric measurements allow us to independently study the real and imaginary parts of the transmittance and, thereby, to test further the dynamic Lifshitz-Kosevich formula \eqref{eq: qc}.

In~Fig.~\ref{fig3}~(b) and (c) we show the magnetic field dependences of the magnitude of the parallel transmittance amplitude $|t_\text{p}|$ and its phase $\varphi$, respectively, measured at 360\,GHz on the GaAs sample. 
Both signals reveal optical SdH oscillations. 
Panels~(d) and (e) show the same data in $1/B$ scale, with the smooth Drude background subtracted. 
Thanks to higher frequency, more nodes are resolved here. 
There is one universal node at $\omega/\omega_c=1$ on both curves, other nodes are tunable and have different  magnetic field positions in the transmittance amplitude and its phase.
The theory curves, shown in red, are calculated  using Eqs.~\eqref{eq: TD} and \eqref{eq: qc}, and they demonstrate that the positions of all nodes are well reproduced using the same effective mass as in Fig.~\ref{fig1}, $m = 0.073\,m_0$. 
It should be mentioned that some deviations between the curves can be due to standing waves in the optical setup, 
see Supplementary Sec.~\ref{SM_Spectrum} bellow, and Ref.~\cite{Savchenko2022} for more details.
Further on, the high-field node position in~Fig.~\ref{fig3}~(b) at $1/B \approx 0.63\,$T$^{-1}$ is poorly fitted. This can be due to the transition to a regime of separated Landau levels which requires next-order expansion terms in Eq.~\eqref{eq: qc}. In theoretical fits, we omitted the CR regions where the condition $\mu (B_\text{CR} - B) \gg \delta$ no longer holds. At the same time, it is seen that the phase shift of the calculated optical SdH oscillations agrees well with the experiment on both sides of the CR, which confirms the node at $\omega=\omega_\text{c}$. 
Overall, the comparison shows that Eq.~\eqref{eq: qc} for the optical SdH oscillations works well, and reproduces  the position of nodes and the phase jumps of the oscillations.

\textit{Summary and outlook.}
The observed SdH oscillations in transmittance are as fundamental as their static counterpart that provides a powerful tool to characterize 2DES. They are formed in the optical response of the system irrespective of the type of the band dispersion, and their main frequency is precisely determined by the 2DES density.
In contrast to the well known static SdH oscillations, the dynamic oscillations in transmittance have an extra modulation that is controlled by the $\omega/\omega_\text{c}$ ratio in a unique way -- the quantum conductivity correction has an imaginary part that is as essential as the real part. 
There are universal nodes in the transmittance oscillations at integer $\omega/\omega_\text{c}$ that should appear in the same positions in absorption and transmission. There is also a similar number of additional, tunable nodes that appear at different position in the amplitude and phase measurements of the transmission and reflection, while in the absorption they 
translate into the nodes at half-integer $\omega/\omega_\text{c}$~\cite{Fedorych2010}.
The optically tunable nodes in transmittance can be explored using both constant-frequency and time-domain set-ups, and are sensitive to all parameters of the structure allowing to determine these parameters with high accuracy.

We acknowledge the financial support of the Austrian Science Funds (I 3456-N27,  I 5539-N). I.A.D. acknowledges the support from the German Research Foundation (Deutsche Forschungsgemeinschaft) via project DM1-6/1.

	\bibliography{library.bib}

\begin{thebibliography}{35}%
\makeatletter
\providecommand \@ifxundefined [1]{%
 \@ifx{#1\undefined}
}%
\providecommand \@ifnum [1]{%
 \ifnum #1\expandafter \@firstoftwo
 \else \expandafter \@secondoftwo
 \fi
}%
\providecommand \@ifx [1]{%
 \ifx #1\expandafter \@firstoftwo
 \else \expandafter \@secondoftwo
 \fi
}%
\providecommand \natexlab [1]{#1}%
\providecommand \enquote  [1]{``#1''}%
\providecommand \bibnamefont  [1]{#1}%
\providecommand \bibfnamefont [1]{#1}%
\providecommand \citenamefont [1]{#1}%
\providecommand \href@noop [0]{\@secondoftwo}%
\providecommand \href [0]{\begingroup \@sanitize@url \@href}%
\providecommand \@href[1]{\@@startlink{#1}\@@href}%
\providecommand \@@href[1]{\endgroup#1\@@endlink}%
\providecommand \@sanitize@url [0]{\catcode `\\12\catcode `\$12\catcode
  `\&12\catcode `\#12\catcode `\^12\catcode `\_12\catcode `\%12\relax}%
\providecommand \@@startlink[1]{}%
\providecommand \@@endlink[0]{}%
\providecommand \url  [0]{\begingroup\@sanitize@url \@url }%
\providecommand \@url [1]{\endgroup\@href {#1}{\urlprefix }}%
\providecommand \urlprefix  [0]{URL }%
\providecommand \Eprint [0]{\href }%
\providecommand \doibase [0]{https://doi.org/}%
\providecommand \selectlanguage [0]{\@gobble}%
\providecommand \bibinfo  [0]{\@secondoftwo}%
\providecommand \bibfield  [0]{\@secondoftwo}%
\providecommand \translation [1]{[#1]}%
\providecommand \BibitemOpen [0]{}%
\providecommand \bibitemStop [0]{}%
\providecommand \bibitemNoStop [0]{.\EOS\space}%
\providecommand \EOS [0]{\spacefactor3000\relax}%
\providecommand \BibitemShut  [1]{\csname bibitem#1\endcsname}%
\let\auto@bib@innerbib\@empty
\bibitem [{\citenamefont {Ando}\ \emph {et~al.}(1982)\citenamefont {Ando},
  \citenamefont {Fowler},\ and\ \citenamefont {Stern}}]{Ando1982}%
  \BibitemOpen
  \bibfield  {author} {\bibinfo {author} {\bibfnamefont {T.}~\bibnamefont
  {Ando}}, \bibinfo {author} {\bibfnamefont {A.~B.}\ \bibnamefont {Fowler}},\
  and\ \bibinfo {author} {\bibfnamefont {F.}~\bibnamefont {Stern}},\ }\bibfield
   {title} {\bibinfo {title} {{Electronic properties of two-dimensional
  systems}},\ }\href {https://doi.org/10.1103/RevModPhys.54.437} {\bibfield
  {journal} {\bibinfo  {journal} {Rev. Mod. Phys.}\ }\textbf {\bibinfo {volume}
  {54}},\ \bibinfo {pages} {437} (\bibinfo {year} {1982})}\BibitemShut
  {NoStop}%
\bibitem [{\citenamefont {Abstreiter}\ \emph {et~al.}(1976)\citenamefont
  {Abstreiter}, \citenamefont {Kotthaus}, \citenamefont {Koch},\ and\
  \citenamefont {Dord}}]{Abstreiter1976}%
  \BibitemOpen
  \bibfield  {author} {\bibinfo {author} {\bibfnamefont {G.}~\bibnamefont
  {Abstreiter}}, \bibinfo {author} {\bibfnamefont {J.~P.}\ \bibnamefont
  {Kotthaus}}, \bibinfo {author} {\bibfnamefont {J.~F.}\ \bibnamefont {Koch}},\
  and\ \bibinfo {author} {\bibfnamefont {G.}~\bibnamefont {Dord}},\ }\bibfield
  {title} {\bibinfo {title} {{Cyclotron resonance of electrons in surface
  space-charge layers on silicon}},\ }\href
  {https://doi.org/https://doi.org/10.1103/PhysRevB.14.2480} {\bibfield
  {journal} {\bibinfo  {journal} {Phys. Rev. B}\ }\textbf {\bibinfo {volume}
  {14}},\ \bibinfo {pages} {2480} (\bibinfo {year} {1976})}\BibitemShut
  {NoStop}%
\bibitem [{\citenamefont {Fedorych}\ \emph {et~al.}(2010)\citenamefont
  {Fedorych}, \citenamefont {Potemski}, \citenamefont {Studenikin},
  \citenamefont {Gupta}, \citenamefont {Wasilewski},\ and\ \citenamefont
  {Dmitriev}}]{Fedorych2010}%
  \BibitemOpen
  \bibfield  {author} {\bibinfo {author} {\bibfnamefont {O.~M.}\ \bibnamefont
  {Fedorych}}, \bibinfo {author} {\bibfnamefont {M.}~\bibnamefont {Potemski}},
  \bibinfo {author} {\bibfnamefont {S.~A.}\ \bibnamefont {Studenikin}},
  \bibinfo {author} {\bibfnamefont {J.~A.}\ \bibnamefont {Gupta}}, \bibinfo
  {author} {\bibfnamefont {Z.~R.}\ \bibnamefont {Wasilewski}},\ and\ \bibinfo
  {author} {\bibfnamefont {I.~A.}\ \bibnamefont {Dmitriev}},\ }\bibfield
  {title} {\bibinfo {title} {{Quantum oscillations in the microwave
  magnetoabsorption of a two-dimensional electron gas}},\ }\href
  {https://doi.org/10.1103/PhysRevB.81.201302} {\bibfield  {journal} {\bibinfo
  {journal} {Phys. Rev. B}\ }\textbf {\bibinfo {volume} {81}},\ \bibinfo
  {pages} {201302} (\bibinfo {year} {2010})}\BibitemShut {NoStop}%
\bibitem [{\citenamefont {Shuvaev}\ \emph {et~al.}(2013)\citenamefont
  {Shuvaev}, \citenamefont {Astakhov}, \citenamefont {Tkachov}, \citenamefont
  {Br{\"{u}}ne}, \citenamefont {Buhmann}, \citenamefont {Molenkamp},\ and\
  \citenamefont {Pimenov}}]{Shuvaev2013}%
  \BibitemOpen
  \bibfield  {author} {\bibinfo {author} {\bibfnamefont {A.~M.}\ \bibnamefont
  {Shuvaev}}, \bibinfo {author} {\bibfnamefont {G.~V.}\ \bibnamefont
  {Astakhov}}, \bibinfo {author} {\bibfnamefont {G.}~\bibnamefont {Tkachov}},
  \bibinfo {author} {\bibfnamefont {C.}~\bibnamefont {Br{\"{u}}ne}}, \bibinfo
  {author} {\bibfnamefont {H.}~\bibnamefont {Buhmann}}, \bibinfo {author}
  {\bibfnamefont {L.~W.}\ \bibnamefont {Molenkamp}},\ and\ \bibinfo {author}
  {\bibfnamefont {A.}~\bibnamefont {Pimenov}},\ }\bibfield  {title} {\bibinfo
  {title} {{Terahertz quantum Hall effect of Dirac fermions in a topological
  insulator}},\ }\href {https://doi.org/10.1103/PhysRevB.87.121104} {\bibfield
  {journal} {\bibinfo  {journal} {Phys. Rev. B}\ }\textbf {\bibinfo {volume}
  {87}},\ \bibinfo {pages} {121104(R)} (\bibinfo {year} {2013})}\BibitemShut
  {NoStop}%
\bibitem [{\citenamefont {Dziom}\ \emph {et~al.}(2019)\citenamefont {Dziom},
  \citenamefont {Shuvaev}, \citenamefont {Shchepetilnikov}, \citenamefont
  {MacFarland}, \citenamefont {Strasser},\ and\ \citenamefont
  {Pimenov}}]{Dziom2019}%
  \BibitemOpen
  \bibfield  {author} {\bibinfo {author} {\bibfnamefont {V.}~\bibnamefont
  {Dziom}}, \bibinfo {author} {\bibfnamefont {A.}~\bibnamefont {Shuvaev}},
  \bibinfo {author} {\bibfnamefont {A.~V.}\ \bibnamefont {Shchepetilnikov}},
  \bibinfo {author} {\bibfnamefont {D.}~\bibnamefont {MacFarland}}, \bibinfo
  {author} {\bibfnamefont {G.}~\bibnamefont {Strasser}},\ and\ \bibinfo
  {author} {\bibfnamefont {A.}~\bibnamefont {Pimenov}},\ }\bibfield  {title}
  {\bibinfo {title} {{High-frequency breakdown of the integer quantum Hall
  effect in GaAs/AlGaAs heterojunctions}},\ }\href
  {https://doi.org/10.1103/PhysRevB.99.045305} {\bibfield  {journal} {\bibinfo
  {journal} {Phys. Rev. B}\ }\textbf {\bibinfo {volume} {99}},\ \bibinfo
  {pages} {045305} (\bibinfo {year} {2019})}\BibitemShut {NoStop}%
\bibitem [{\citenamefont {Ando}(1975)}]{Ando1975}%
  \BibitemOpen
  \bibfield  {author} {\bibinfo {author} {\bibfnamefont {T.}~\bibnamefont
  {Ando}},\ }\bibfield  {title} {\bibinfo {title} {{Theory of Cyclotron
  Resonance Lineshape in a Two-Dimensional Electron System}},\ }\href
  {https://doi.org/10.1143/JPSJ.38.989} {\bibfield  {journal} {\bibinfo
  {journal} {J. Phys. Soc. Japan}\ }\textbf {\bibinfo {volume} {38}},\ \bibinfo
  {pages} {989} (\bibinfo {year} {1975})}\BibitemShut {NoStop}%
\bibitem [{\citenamefont {Dmitriev}\ \emph {et~al.}(2003)\citenamefont
  {Dmitriev}, \citenamefont {Mirlin},\ and\ \citenamefont
  {Polyakov}}]{Dmitriev2003}%
  \BibitemOpen
  \bibfield  {author} {\bibinfo {author} {\bibfnamefont {I.~A.}\ \bibnamefont
  {Dmitriev}}, \bibinfo {author} {\bibfnamefont {A.~D.}\ \bibnamefont
  {Mirlin}},\ and\ \bibinfo {author} {\bibfnamefont {D.~G.}\ \bibnamefont
  {Polyakov}},\ }\bibfield  {title} {\bibinfo {title} {{Cyclotron-Resonance
  Harmonics in the ac Response of a 2D Electron Gas with Smooth Disorder}},\
  }\href {https://doi.org/10.1103/PhysRevLett.91.226802} {\bibfield  {journal}
  {\bibinfo  {journal} {Phys. Rev. Lett.}\ }\textbf {\bibinfo {volume} {91}},\
  \bibinfo {pages} {226802} (\bibinfo {year} {2003})}\BibitemShut {NoStop}%
\bibitem [{\citenamefont {Raichev}(2008)}]{Raichev2008}%
  \BibitemOpen
  \bibfield  {author} {\bibinfo {author} {\bibfnamefont {O.~E.}\ \bibnamefont
  {Raichev}},\ }\bibfield  {title} {\bibinfo {title} {{Magnetic oscillations of
  resistivity and absorption of radiation in quantum wells with two populated
  subbands}},\ }\href {https://doi.org/10.1103/PhysRevB.78.125304} {\bibfield
  {journal} {\bibinfo  {journal} {Phys. Rev. B}\ }\textbf {\bibinfo {volume}
  {78}},\ \bibinfo {pages} {125304} (\bibinfo {year} {2008})}\BibitemShut
  {NoStop}%
\bibitem [{\citenamefont {Savchenko}\ \emph {et~al.}(2022)\citenamefont
  {Savchenko}, \citenamefont {Shuvaev}, \citenamefont {Dmitriev}, \citenamefont
  {Ganichev}, \citenamefont {Kvon},\ and\ \citenamefont
  {Pimenov}}]{Savchenko2022}%
  \BibitemOpen
  \bibfield  {author} {\bibinfo {author} {\bibfnamefont {M.~L.}\ \bibnamefont
  {Savchenko}}, \bibinfo {author} {\bibfnamefont {A.}~\bibnamefont {Shuvaev}},
  \bibinfo {author} {\bibfnamefont {I.~A.}\ \bibnamefont {Dmitriev}}, \bibinfo
  {author} {\bibfnamefont {S.~D.}\ \bibnamefont {Ganichev}}, \bibinfo {author}
  {\bibfnamefont {Z.~D.}\ \bibnamefont {Kvon}},\ and\ \bibinfo {author}
  {\bibfnamefont {A.}~\bibnamefont {Pimenov}},\ }\bibfield  {title} {\bibinfo
  {title} {{Demonstration of high sensitivity of microwave-induced resistance
  oscillations to circular polarization}},\ }\href
  {https://doi.org/https://doi.org/10.1103/PhysRevB.106.L161408} {\bibfield
  {journal} {\bibinfo  {journal} {Phys. Rev. B}\ }\textbf {\bibinfo {volume}
  {106}},\ \bibinfo {pages} {L161408} (\bibinfo {year} {2022})}\BibitemShut
  {NoStop}%
\bibitem [{\citenamefont {Shoenberg}(1984)}]{Shoenberg1984}%
  \BibitemOpen
  \bibfield  {author} {\bibinfo {author} {\bibfnamefont {D.}~\bibnamefont
  {Shoenberg}},\ }\bibfield  {title} {\bibinfo {title} {{Magnetic Oscillations
  in Metals}},\ }\bibfield  {journal} {\bibinfo  {journal} {Cambridge Univ.
  Press}\ }\href {https://doi.org/https://doi.org/10.1017/CBO9780511897870}
  {https://doi.org/10.1017/CBO9780511897870} (\bibinfo {year}
  {1984})\BibitemShut {NoStop}%
\bibitem [{\citenamefont {Dmitriev}\ \emph {et~al.}(2012)\citenamefont
  {Dmitriev}, \citenamefont {Mirlin}, \citenamefont {Polyakov},\ and\
  \citenamefont {Zudov}}]{Dmitriev2012}%
  \BibitemOpen
  \bibfield  {author} {\bibinfo {author} {\bibfnamefont {I.~A.}\ \bibnamefont
  {Dmitriev}}, \bibinfo {author} {\bibfnamefont {A.~D.}\ \bibnamefont
  {Mirlin}}, \bibinfo {author} {\bibfnamefont {D.~G.}\ \bibnamefont
  {Polyakov}},\ and\ \bibinfo {author} {\bibfnamefont {M.~A.}\ \bibnamefont
  {Zudov}},\ }\bibfield  {title} {\bibinfo {title} {{Nonequilibrium phenomena
  in high Landau levels}},\ }\href {https://doi.org/10.1103/RevModPhys.84.1709}
  {\bibfield  {journal} {\bibinfo  {journal} {Rev. Mod. Phys.}\ }\textbf
  {\bibinfo {volume} {84}},\ \bibinfo {pages} {1709} (\bibinfo {year}
  {2012})}\BibitemShut {NoStop}%
\bibitem [{\citenamefont {Baba}\ \emph {et~al.}(1983)\citenamefont {Baba},
  \citenamefont {Mizutani},\ and\ \citenamefont {Ogawa}}]{Baba1983}%
  \BibitemOpen
  \bibfield  {author} {\bibinfo {author} {\bibfnamefont {T.}~\bibnamefont
  {Baba}}, \bibinfo {author} {\bibfnamefont {T.}~\bibnamefont {Mizutani}},\
  and\ \bibinfo {author} {\bibfnamefont {M.}~\bibnamefont {Ogawa}},\ }\bibfield
   {title} {\bibinfo {title} {{Elimination of Persistent Photoconductivity and
  Improvement in Si Activation Coefficient by Al Spatial Separation from Ga and
  Si in Al-Ga-As:Si Solid System -- a Novel Short Period AlAs/n-GaAs
  Superlattice --}},\ }\href {https://doi.org/10.1143/JJAP.22.L627} {\bibfield
  {journal} {\bibinfo  {journal} {Jpn. J. Appl. Phys.}\ }\textbf {\bibinfo
  {volume} {22}},\ \bibinfo {pages} {L627} (\bibinfo {year}
  {1983})}\BibitemShut {NoStop}%
\bibitem [{\citenamefont {Friedland}\ \emph {et~al.}(1996)\citenamefont
  {Friedland}, \citenamefont {Hey}, \citenamefont {Kostial}, \citenamefont
  {Klann},\ and\ \citenamefont {Ploog}}]{Friedland1996}%
  \BibitemOpen
  \bibfield  {author} {\bibinfo {author} {\bibfnamefont {K.-J.}\ \bibnamefont
  {Friedland}}, \bibinfo {author} {\bibfnamefont {R.}~\bibnamefont {Hey}},
  \bibinfo {author} {\bibfnamefont {H.}~\bibnamefont {Kostial}}, \bibinfo
  {author} {\bibfnamefont {R.}~\bibnamefont {Klann}},\ and\ \bibinfo {author}
  {\bibfnamefont {K.}~\bibnamefont {Ploog}},\ }\bibfield  {title} {\bibinfo
  {title} {{New Concept for the Reduction of Impurity Scattering in Remotely
  Doped GaAs Quantum Wells}},\ }\href
  {https://doi.org/10.1103/PhysRevLett.77.4616} {\bibfield  {journal} {\bibinfo
   {journal} {Phys. Rev. Lett.}\ }\textbf {\bibinfo {volume} {77}},\ \bibinfo
  {pages} {4616} (\bibinfo {year} {1996})}\BibitemShut {NoStop}%
\bibitem [{\citenamefont {Umansky}\ \emph {et~al.}(2009)\citenamefont
  {Umansky}, \citenamefont {Heiblum}, \citenamefont {Levinson}, \citenamefont
  {Smet}, \citenamefont {N{\"{u}}bler},\ and\ \citenamefont
  {Dolev}}]{Umansky2009}%
  \BibitemOpen
  \bibfield  {author} {\bibinfo {author} {\bibfnamefont {V.}~\bibnamefont
  {Umansky}}, \bibinfo {author} {\bibfnamefont {M.}~\bibnamefont {Heiblum}},
  \bibinfo {author} {\bibfnamefont {Y.}~\bibnamefont {Levinson}}, \bibinfo
  {author} {\bibfnamefont {J.}~\bibnamefont {Smet}}, \bibinfo {author}
  {\bibfnamefont {J.}~\bibnamefont {N{\"{u}}bler}},\ and\ \bibinfo {author}
  {\bibfnamefont {M.}~\bibnamefont {Dolev}},\ }\bibfield  {title} {\bibinfo
  {title} {{MBE growth of ultra-low disorder 2DEG with mobility exceeding
  35×106cm2/Vs}},\ }\href {https://doi.org/10.1016/j.jcrysgro.2008.09.151}
  {\bibfield  {journal} {\bibinfo  {journal} {J. Cryst. Growth}\ }\textbf
  {\bibinfo {volume} {311}},\ \bibinfo {pages} {1658} (\bibinfo {year}
  {2009})}\BibitemShut {NoStop}%
\bibitem [{\citenamefont {Manfra}(2014)}]{Manfra2014}%
  \BibitemOpen
  \bibfield  {author} {\bibinfo {author} {\bibfnamefont {M.~J.}\ \bibnamefont
  {Manfra}},\ }\bibfield  {title} {\bibinfo {title} {{Molecular Beam Epitaxy of
  Ultra-High-Quality AlGaAs/GaAs Heterostructures: Enabling Physics in
  Low-Dimensional Electronic Systems}},\ }\href
  {https://doi.org/10.1146/annurev-conmatphys-031113-133905} {\bibfield
  {journal} {\bibinfo  {journal} {Annu. Rev. Condens. Matter Phys.}\ }\textbf
  {\bibinfo {volume} {5}},\ \bibinfo {pages} {347} (\bibinfo {year}
  {2014})}\BibitemShut {NoStop}%
\bibitem [{\citenamefont {Kvon}\ \emph {et~al.}(2011)\citenamefont {Kvon},
  \citenamefont {Danilov}, \citenamefont {Kozlov}, \citenamefont {Zoth},
  \citenamefont {Mikhailov}, \citenamefont {Dvoretskii},\ and\ \citenamefont
  {Ganichev}}]{Kvon2011}%
  \BibitemOpen
  \bibfield  {author} {\bibinfo {author} {\bibfnamefont {Z.~D.}\ \bibnamefont
  {Kvon}}, \bibinfo {author} {\bibfnamefont {S.~N.}\ \bibnamefont {Danilov}},
  \bibinfo {author} {\bibfnamefont {D.~A.}\ \bibnamefont {Kozlov}}, \bibinfo
  {author} {\bibfnamefont {C.}~\bibnamefont {Zoth}}, \bibinfo {author}
  {\bibfnamefont {N.~N.}\ \bibnamefont {Mikhailov}}, \bibinfo {author}
  {\bibfnamefont {S.~A.}\ \bibnamefont {Dvoretskii}},\ and\ \bibinfo {author}
  {\bibfnamefont {S.~D.}\ \bibnamefont {Ganichev}},\ }\bibfield  {title}
  {\bibinfo {title} {{Cyclotron Resonance of Dirac Ferions in HgTe Quantum
  Wells}},\ }\href {https://doi.org/10.1134/S002136401123007X} {\bibfield
  {journal} {\bibinfo  {journal} {JETP Lett.}\ }\textbf {\bibinfo {volume}
  {94}},\ \bibinfo {pages} {816} (\bibinfo {year} {2011})}\BibitemShut
  {NoStop}%
\bibitem [{\citenamefont {Shuvaev}\ \emph {et~al.}(2022)\citenamefont
  {Shuvaev}, \citenamefont {Dziom}, \citenamefont {Gospodari{\v{c}}},
  \citenamefont {Novik}, \citenamefont {Dobretsova}, \citenamefont {Mikhailov},
  \citenamefont {Kvon},\ and\ \citenamefont {Pimenov}}]{Shuvaev2022}%
  \BibitemOpen
  \bibfield  {author} {\bibinfo {author} {\bibfnamefont {A.}~\bibnamefont
  {Shuvaev}}, \bibinfo {author} {\bibfnamefont {V.}~\bibnamefont {Dziom}},
  \bibinfo {author} {\bibfnamefont {J.}~\bibnamefont {Gospodari{\v{c}}}},
  \bibinfo {author} {\bibfnamefont {E.~G.}\ \bibnamefont {Novik}}, \bibinfo
  {author} {\bibfnamefont {A.~A.}\ \bibnamefont {Dobretsova}}, \bibinfo
  {author} {\bibfnamefont {N.~N.}\ \bibnamefont {Mikhailov}}, \bibinfo {author}
  {\bibfnamefont {Z.~D.}\ \bibnamefont {Kvon}},\ and\ \bibinfo {author}
  {\bibfnamefont {A.}~\bibnamefont {Pimenov}},\ }\bibfield  {title} {\bibinfo
  {title} {{Band Structure Near the Dirac Point in HgTe Quantum Wells with
  Critical Thickness}},\ }\href {https://doi.org/10.3390/nano12142492}
  {\bibfield  {journal} {\bibinfo  {journal} {Nanomaterials}\ }\textbf
  {\bibinfo {volume} {12}},\ \bibinfo {pages} {2492} (\bibinfo {year}
  {2022})}\BibitemShut {NoStop}%
\bibitem [{\citenamefont {Hopkins}\ \emph {et~al.}(1987)\citenamefont
  {Hopkins}, \citenamefont {Nicholas}, \citenamefont {Brummell},\ and\
  \citenamefont {Harris}}]{Hopkins1987}%
  \BibitemOpen
  \bibfield  {author} {\bibinfo {author} {\bibfnamefont {M.~A.}\ \bibnamefont
  {Hopkins}}, \bibinfo {author} {\bibfnamefont {R.~J.}\ \bibnamefont
  {Nicholas}}, \bibinfo {author} {\bibfnamefont {M.~A.}\ \bibnamefont
  {Brummell}},\ and\ \bibinfo {author} {\bibfnamefont {J.~J.}\ \bibnamefont
  {Harris}},\ }\bibfield  {title} {\bibinfo {title} {{Cyclotron-resonance study
  of nonparabolicity and screening in GaAs-GaAlAs heterojunctions}},\ }\href
  {https://doi.org/https://doi.org/10.1103/PhysRevB.36.4789} {\bibfield
  {journal} {\bibinfo  {journal} {Phys. Rev. B}\ }\textbf {\bibinfo {volume}
  {36}},\ \bibinfo {pages} {4789} (\bibinfo {year} {1987})}\BibitemShut
  {NoStop}%
\bibitem [{\citenamefont {Kukushkin}\ and\ \citenamefont
  {Schmult}(2015)}]{Kukushkin2015}%
  \BibitemOpen
  \bibfield  {author} {\bibinfo {author} {\bibfnamefont {I.~V.}\ \bibnamefont
  {Kukushkin}}\ and\ \bibinfo {author} {\bibfnamefont {S.}~\bibnamefont
  {Schmult}},\ }\bibfield  {title} {\bibinfo {title} {{Fermi liquid effects and
  quasiparticle mass renormalization in a system of two-dimensional electrons
  with strong interaction}},\ }\href
  {https://doi.org/10.1134/S0021364015100082} {\bibfield  {journal} {\bibinfo
  {journal} {JETP Lett.}\ }\textbf {\bibinfo {volume} {101}},\ \bibinfo {pages}
  {693} (\bibinfo {year} {2015})}\BibitemShut {NoStop}%
\bibitem [{\citenamefont {Das}\ \emph {et~al.}(1989)\citenamefont {Das},
  \citenamefont {Miller}, \citenamefont {Datta}, \citenamefont {Reifenberger},
  \citenamefont {Hong}, \citenamefont {Bhattacharya}, \citenamefont {Singh},\
  and\ \citenamefont {Jaffe}}]{Das1989}%
  \BibitemOpen
  \bibfield  {author} {\bibinfo {author} {\bibfnamefont {B.}~\bibnamefont
  {Das}}, \bibinfo {author} {\bibfnamefont {D.~C.}\ \bibnamefont {Miller}},
  \bibinfo {author} {\bibfnamefont {S.}~\bibnamefont {Datta}}, \bibinfo
  {author} {\bibfnamefont {R.}~\bibnamefont {Reifenberger}}, \bibinfo {author}
  {\bibfnamefont {W.~P.}\ \bibnamefont {Hong}}, \bibinfo {author}
  {\bibfnamefont {P.~K.}\ \bibnamefont {Bhattacharya}}, \bibinfo {author}
  {\bibfnamefont {J.}~\bibnamefont {Singh}},\ and\ \bibinfo {author}
  {\bibfnamefont {M.}~\bibnamefont {Jaffe}},\ }\bibfield  {title} {\bibinfo
  {title} {{Evidence for spin splitting in InxGa1-xAs/In0.52Al0.48As
  heterostructures as B -> 0}},\ }\href
  {https://doi.org/10.1103/PhysRevB.39.1411} {\bibfield  {journal} {\bibinfo
  {journal} {Phys. Rev. B}\ }\textbf {\bibinfo {volume} {39}},\ \bibinfo
  {pages} {1411} (\bibinfo {year} {1989})}\BibitemShut {NoStop}%
\bibitem [{\citenamefont {Hatke}\ \emph {et~al.}(2012)\citenamefont {Hatke},
  \citenamefont {Zudov}, \citenamefont {Pfeiffer},\ and\ \citenamefont
  {West}}]{Hatke2012}%
  \BibitemOpen
  \bibfield  {author} {\bibinfo {author} {\bibfnamefont {A.~T.}\ \bibnamefont
  {Hatke}}, \bibinfo {author} {\bibfnamefont {M.~A.}\ \bibnamefont {Zudov}},
  \bibinfo {author} {\bibfnamefont {L.~N.}\ \bibnamefont {Pfeiffer}},\ and\
  \bibinfo {author} {\bibfnamefont {K.~W.}\ \bibnamefont {West}},\ }\bibfield
  {title} {\bibinfo {title} {{Shubnikov-de Haas oscillations in GaAs quantum
  wells in tilted magnetic fields}},\ }\href
  {https://doi.org/10.1103/PhysRevB.85.241305} {\bibfield  {journal} {\bibinfo
  {journal} {Phys. Rev. B}\ }\textbf {\bibinfo {volume} {85}},\ \bibinfo
  {pages} {241305} (\bibinfo {year} {2012})}\BibitemShut {NoStop}%
\bibitem [{\citenamefont {Volkov}\ and\ \citenamefont
  {Pankratov}(1985)}]{Volkov1985}%
  \BibitemOpen
  \bibfield  {author} {\bibinfo {author} {\bibfnamefont {B.~A.}\ \bibnamefont
  {Volkov}}\ and\ \bibinfo {author} {\bibfnamefont {O.~A.}\ \bibnamefont
  {Pankratov}},\ }\bibfield  {title} {\bibinfo {title} {{Two-dimensional
  massless electrons in an inverted contact}},\ }\href
  {http://www.jetpletters.ac.ru/ps/1420/article_21570.shtml} {\bibfield
  {journal} {\bibinfo  {journal} {JETP Lett.}\ }\textbf {\bibinfo {volume}
  {42}},\ \bibinfo {pages} {178} (\bibinfo {year} {1985})}\BibitemShut
  {NoStop}%
\bibitem [{\citenamefont {Buttner}\ \emph {et~al.}(2011)\citenamefont
  {Buttner}, \citenamefont {Liu}, \citenamefont {Tkachov}, \citenamefont
  {Novik}, \citenamefont {Brune}, \citenamefont {Buhmann}, \citenamefont
  {Hankiewicz}, \citenamefont {Recher}, \citenamefont {Trauzettel},
  \citenamefont {Zhang},\ and\ \citenamefont {Molenkamp}}]{Buttner2011}%
  \BibitemOpen
  \bibfield  {author} {\bibinfo {author} {\bibfnamefont {B.}~\bibnamefont
  {Buttner}}, \bibinfo {author} {\bibfnamefont {C.~X.}\ \bibnamefont {Liu}},
  \bibinfo {author} {\bibfnamefont {G.}~\bibnamefont {Tkachov}}, \bibinfo
  {author} {\bibfnamefont {E.~G.}\ \bibnamefont {Novik}}, \bibinfo {author}
  {\bibfnamefont {C.}~\bibnamefont {Brune}}, \bibinfo {author} {\bibfnamefont
  {H.}~\bibnamefont {Buhmann}}, \bibinfo {author} {\bibfnamefont {E.~M.}\
  \bibnamefont {Hankiewicz}}, \bibinfo {author} {\bibfnamefont
  {P.}~\bibnamefont {Recher}}, \bibinfo {author} {\bibfnamefont
  {B.}~\bibnamefont {Trauzettel}}, \bibinfo {author} {\bibfnamefont {S.~C.}\
  \bibnamefont {Zhang}},\ and\ \bibinfo {author} {\bibfnamefont {L.~W.}\
  \bibnamefont {Molenkamp}},\ }\bibfield  {title} {\bibinfo {title} {{Single
  valley Dirac fermions in zero-gap HgTe quantum wells}},\ }\href
  {https://doi.org/10.1038/nphys1914} {\bibfield  {journal} {\bibinfo
  {journal} {Nat. Phys.}\ }\textbf {\bibinfo {volume} {7}},\ \bibinfo {pages}
  {418} (\bibinfo {year} {2011})}\BibitemShut {NoStop}%
\bibitem [{\citenamefont {Dziom}\ \emph
  {et~al.}(2017{\natexlab{a}})\citenamefont {Dziom}, \citenamefont {Shuvaev},
  \citenamefont {Mikhailov},\ and\ \citenamefont {Pimenov}}]{Dziom2017}%
  \BibitemOpen
  \bibfield  {author} {\bibinfo {author} {\bibfnamefont {V.}~\bibnamefont
  {Dziom}}, \bibinfo {author} {\bibfnamefont {A.}~\bibnamefont {Shuvaev}},
  \bibinfo {author} {\bibfnamefont {N.~N.}\ \bibnamefont {Mikhailov}},\ and\
  \bibinfo {author} {\bibfnamefont {A.}~\bibnamefont {Pimenov}},\ }\bibfield
  {title} {\bibinfo {title} {{Terahertz properties of Dirac fermions in HgTe
  films with optical doping}},\ }\href
  {https://doi.org/10.1088/2053-1583/aa5cd7} {\bibfield  {journal} {\bibinfo
  {journal} {2D Mater.}\ }\textbf {\bibinfo {volume} {4}},\ \bibinfo {pages}
  {024005} (\bibinfo {year} {2017}{\natexlab{a}})}\BibitemShut {NoStop}%
\bibitem [{\citenamefont {Ikonnikov}\ \emph {et~al.}(2011)\citenamefont
  {Ikonnikov}, \citenamefont {Zholudev}, \citenamefont {Spirin}, \citenamefont
  {Lastovkin}, \citenamefont {Maremyanin}, \citenamefont {Aleshkin},
  \citenamefont {Gavrilenko}, \citenamefont {Drachenko}, \citenamefont {Helm},
  \citenamefont {Wosnitza}, \citenamefont {Goiran}, \citenamefont {Mikhailov},
  \citenamefont {Dvoretskii}, \citenamefont {Teppe}, \citenamefont {Diakonova},
  \citenamefont {Consejo}, \citenamefont {Chenaud},\ and\ \citenamefont
  {Knap}}]{Ikonnikov2011}%
  \BibitemOpen
  \bibfield  {author} {\bibinfo {author} {\bibfnamefont {A.~V.}\ \bibnamefont
  {Ikonnikov}}, \bibinfo {author} {\bibfnamefont {M.~S.}\ \bibnamefont
  {Zholudev}}, \bibinfo {author} {\bibfnamefont {K.~E.}\ \bibnamefont
  {Spirin}}, \bibinfo {author} {\bibfnamefont {A.~A.}\ \bibnamefont
  {Lastovkin}}, \bibinfo {author} {\bibfnamefont {K.~V.}\ \bibnamefont
  {Maremyanin}}, \bibinfo {author} {\bibfnamefont {V.~Y.}\ \bibnamefont
  {Aleshkin}}, \bibinfo {author} {\bibfnamefont {V.~I.}\ \bibnamefont
  {Gavrilenko}}, \bibinfo {author} {\bibfnamefont {O.}~\bibnamefont
  {Drachenko}}, \bibinfo {author} {\bibfnamefont {M.}~\bibnamefont {Helm}},
  \bibinfo {author} {\bibfnamefont {J.}~\bibnamefont {Wosnitza}}, \bibinfo
  {author} {\bibfnamefont {M.}~\bibnamefont {Goiran}}, \bibinfo {author}
  {\bibfnamefont {N.~N.}\ \bibnamefont {Mikhailov}}, \bibinfo {author}
  {\bibfnamefont {S.~A.}\ \bibnamefont {Dvoretskii}}, \bibinfo {author}
  {\bibfnamefont {F.}~\bibnamefont {Teppe}}, \bibinfo {author} {\bibfnamefont
  {N.}~\bibnamefont {Diakonova}}, \bibinfo {author} {\bibfnamefont
  {C.}~\bibnamefont {Consejo}}, \bibinfo {author} {\bibfnamefont
  {B.}~\bibnamefont {Chenaud}},\ and\ \bibinfo {author} {\bibfnamefont
  {W.}~\bibnamefont {Knap}},\ }\bibfield  {title} {\bibinfo {title} {{Cyclotron
  resonance and interband optical transitions in HgTe/CdTe(0 1 3) quantum well
  heterostructures}},\ }\href {https://doi.org/10.1088/0268-1242/26/12/125011}
  {\bibfield  {journal} {\bibinfo  {journal} {Semicond. Sci. Technol.}\
  }\textbf {\bibinfo {volume} {26}},\ \bibinfo {pages} {125011} (\bibinfo
  {year} {2011})}\BibitemShut {NoStop}%
\bibitem [{\citenamefont {Minkov}\ \emph {et~al.}(2020)\citenamefont {Minkov},
  \citenamefont {Aleshkin}, \citenamefont {Rut}, \citenamefont {Sherstobitov},
  \citenamefont {Germanenko}, \citenamefont {Dvoretski},\ and\ \citenamefont
  {Mikhailov}}]{Minkov2020a}%
  \BibitemOpen
  \bibfield  {author} {\bibinfo {author} {\bibfnamefont {G.~M.}\ \bibnamefont
  {Minkov}}, \bibinfo {author} {\bibfnamefont {V.}~\bibnamefont {Aleshkin}},
  \bibinfo {author} {\bibfnamefont {O.}~\bibnamefont {Rut}}, \bibinfo {author}
  {\bibfnamefont {A.}~\bibnamefont {Sherstobitov}}, \bibinfo {author}
  {\bibfnamefont {A.}~\bibnamefont {Germanenko}}, \bibinfo {author}
  {\bibfnamefont {S.}~\bibnamefont {Dvoretski}},\ and\ \bibinfo {author}
  {\bibfnamefont {N.}~\bibnamefont {Mikhailov}},\ }\bibfield  {title} {\bibinfo
  {title} {{Electron mass in a HgTe quantum well: Experiment versus theory}},\
  }\href {https://doi.org/10.1016/j.physe.2019.113742} {\bibfield  {journal}
  {\bibinfo  {journal} {Phys. E Low-dimensional Syst. Nanostructures}\ }\textbf
  {\bibinfo {volume} {116}},\ \bibinfo {pages} {113742} (\bibinfo {year}
  {2020})}\BibitemShut {NoStop}%
\bibitem [{SM_()}]{SM_universal}%
  \BibitemOpen
  \href@noop {} {}\bibinfo {note} {See Supplemental Material at [URL will be
  inserted by publisher] for measurements at other frequencies, details of the
  fitting procedure, and a discussion on the influence of standing
  waves.}\BibitemShut {Stop}%
\bibitem [{\citenamefont {Hatke}\ \emph {et~al.}(2013)\citenamefont {Hatke},
  \citenamefont {Zudov}, \citenamefont {Watson}, \citenamefont {Manfra},
  \citenamefont {Pfeiffer},\ and\ \citenamefont {West}}]{Hatke2013}%
  \BibitemOpen
  \bibfield  {author} {\bibinfo {author} {\bibfnamefont {A.~T.}\ \bibnamefont
  {Hatke}}, \bibinfo {author} {\bibfnamefont {M.~A.}\ \bibnamefont {Zudov}},
  \bibinfo {author} {\bibfnamefont {J.~D.}\ \bibnamefont {Watson}}, \bibinfo
  {author} {\bibfnamefont {M.~J.}\ \bibnamefont {Manfra}}, \bibinfo {author}
  {\bibfnamefont {L.~N.}\ \bibnamefont {Pfeiffer}},\ and\ \bibinfo {author}
  {\bibfnamefont {K.~W.}\ \bibnamefont {West}},\ }\bibfield  {title} {\bibinfo
  {title} {{Evidence for effective mass reduction in GaAs/AlGaAs quantum
  wells}},\ }\href {https://doi.org/10.1103/PhysRevB.87.161307} {\bibfield
  {journal} {\bibinfo  {journal} {Phys. Rev. B}\ }\textbf {\bibinfo {volume}
  {87}},\ \bibinfo {pages} {161307} (\bibinfo {year} {2013})}\BibitemShut
  {NoStop}%
\bibitem [{\citenamefont {Tabrea}\ \emph {et~al.}(2020)\citenamefont {Tabrea},
  \citenamefont {Dmitriev}, \citenamefont {Dorozhkin}, \citenamefont
  {Gorshunov}, \citenamefont {Boris}, \citenamefont {Kozuka},\ and\
  \citenamefont {Tsukazaki}}]{Tabrea2020}%
  \BibitemOpen
  \bibfield  {author} {\bibinfo {author} {\bibfnamefont {D.}~\bibnamefont
  {Tabrea}}, \bibinfo {author} {\bibfnamefont {I.~A.}\ \bibnamefont
  {Dmitriev}}, \bibinfo {author} {\bibfnamefont {S.~I.}\ \bibnamefont
  {Dorozhkin}}, \bibinfo {author} {\bibfnamefont {B.~P.}\ \bibnamefont
  {Gorshunov}}, \bibinfo {author} {\bibfnamefont {A.~V.}\ \bibnamefont
  {Boris}}, \bibinfo {author} {\bibfnamefont {Y.}~\bibnamefont {Kozuka}},\ and\
  \bibinfo {author} {\bibfnamefont {A.}~\bibnamefont {Tsukazaki}},\ }\bibfield
  {title} {\bibinfo {title} {{Microwave response of interacting oxide
  two-dimensional electron systems}},\ }\href
  {https://doi.org/10.1103/PhysRevB.102.115432} {\bibfield  {journal} {\bibinfo
   {journal} {Phys. Rev. B}\ }\textbf {\bibinfo {volume} {102}},\ \bibinfo
  {pages} {115432} (\bibinfo {year} {2020})}\BibitemShut {NoStop}%
\bibitem [{\citenamefont {Savchenko}\ \emph {et~al.}(2021)\citenamefont
  {Savchenko}, \citenamefont {Shuvaev}, \citenamefont {Dmitriev}, \citenamefont
  {Bykov}, \citenamefont {Bakarov}, \citenamefont {Kvon},\ and\ \citenamefont
  {Pimenov}}]{Savchenko2020b}%
  \BibitemOpen
  \bibfield  {author} {\bibinfo {author} {\bibfnamefont {M.~L.}\ \bibnamefont
  {Savchenko}}, \bibinfo {author} {\bibfnamefont {A.}~\bibnamefont {Shuvaev}},
  \bibinfo {author} {\bibfnamefont {I.~A.}\ \bibnamefont {Dmitriev}}, \bibinfo
  {author} {\bibfnamefont {A.~A.}\ \bibnamefont {Bykov}}, \bibinfo {author}
  {\bibfnamefont {A.~K.}\ \bibnamefont {Bakarov}}, \bibinfo {author}
  {\bibfnamefont {Z.~D.}\ \bibnamefont {Kvon}},\ and\ \bibinfo {author}
  {\bibfnamefont {A.}~\bibnamefont {Pimenov}},\ }\bibfield  {title} {\bibinfo
  {title} {{High harmonics of the cyclotron resonance in microwave transmission
  of a high-mobility two-dimensional electron system}},\ }\href
  {https://doi.org/https://doi.org/10.1103/PhysRevResearch.3.L012013}
  {\bibfield  {journal} {\bibinfo  {journal} {Phys. Rev. Res.}\ }\textbf
  {\bibinfo {volume} {3}},\ \bibinfo {pages} {L012013} (\bibinfo {year}
  {2021})},\ \Eprint {https://arxiv.org/abs/2008.11114} {arXiv:2008.11114}
  \BibitemShut {NoStop}%
\bibitem [{\citenamefont {Volkov}\ \emph {et~al.}(1985)\citenamefont {Volkov},
  \citenamefont {Kozlov}, \citenamefont {Lebedev},\ and\ \citenamefont
  {Prokhorov}}]{Volkov1985a}%
  \BibitemOpen
  \bibfield  {author} {\bibinfo {author} {\bibfnamefont {A.~A.}\ \bibnamefont
  {Volkov}}, \bibinfo {author} {\bibfnamefont {G.~V.}\ \bibnamefont {Kozlov}},
  \bibinfo {author} {\bibfnamefont {S.~P.}\ \bibnamefont {Lebedev}},\ and\
  \bibinfo {author} {\bibfnamefont {A.~M.}\ \bibnamefont {Prokhorov}},\
  }\bibfield  {title} {\bibinfo {title} {{Dielectric measurements in the
  submillimeter wavelength region}},\ }\href@noop {} {\bibfield  {journal}
  {\bibinfo  {journal} {Infrared Phys.}\ }\textbf {\bibinfo {volume} {25}},\
  \bibinfo {pages} {369} (\bibinfo {year} {1985})}\BibitemShut {NoStop}%
\bibitem [{\citenamefont {Shuvaev}\ \emph {et~al.}(2012)\citenamefont
  {Shuvaev}, \citenamefont {Astakhov}, \citenamefont {Br{\"{u}}ne},
  \citenamefont {Buhmann}, \citenamefont {Molenkamp},\ and\ \citenamefont
  {Pimenov}}]{Shuvaev2012}%
  \BibitemOpen
  \bibfield  {author} {\bibinfo {author} {\bibfnamefont {A.~M.}\ \bibnamefont
  {Shuvaev}}, \bibinfo {author} {\bibfnamefont {G.~V.}\ \bibnamefont
  {Astakhov}}, \bibinfo {author} {\bibfnamefont {C.}~\bibnamefont
  {Br{\"{u}}ne}}, \bibinfo {author} {\bibfnamefont {H.}~\bibnamefont
  {Buhmann}}, \bibinfo {author} {\bibfnamefont {L.~W.}\ \bibnamefont
  {Molenkamp}},\ and\ \bibinfo {author} {\bibfnamefont {A.}~\bibnamefont
  {Pimenov}},\ }\bibfield  {title} {\bibinfo {title} {{Terahertz
  magneto-optical spectroscopy in HgTe thin films}},\ }\href
  {https://doi.org/https://doi.org/10.1088/0268-1242/27/12/124004} {\bibfield
  {journal} {\bibinfo  {journal} {Semicond. Sci. Technol.}\ }\textbf {\bibinfo
  {volume} {27}},\ \bibinfo {pages} {124004} (\bibinfo {year}
  {2012})}\BibitemShut {NoStop}%
\bibitem [{\citenamefont {Dziom}\ \emph
  {et~al.}(2017{\natexlab{b}})\citenamefont {Dziom}, \citenamefont {Shuvaev},
  \citenamefont {Pimenov}, \citenamefont {Astakhov}, \citenamefont {Ames},
  \citenamefont {Bendias}, \citenamefont {B{\"{o}}ttcher}, \citenamefont
  {Tkachov}, \citenamefont {Hankiewicz}, \citenamefont {Br{\"{u}}ne},
  \citenamefont {Buhmann},\ and\ \citenamefont {Molenkamp}}]{Dziom2017a}%
  \BibitemOpen
  \bibfield  {author} {\bibinfo {author} {\bibfnamefont {V.}~\bibnamefont
  {Dziom}}, \bibinfo {author} {\bibfnamefont {A.}~\bibnamefont {Shuvaev}},
  \bibinfo {author} {\bibfnamefont {A.}~\bibnamefont {Pimenov}}, \bibinfo
  {author} {\bibfnamefont {G.~V.}\ \bibnamefont {Astakhov}}, \bibinfo {author}
  {\bibfnamefont {C.}~\bibnamefont {Ames}}, \bibinfo {author} {\bibfnamefont
  {K.}~\bibnamefont {Bendias}}, \bibinfo {author} {\bibfnamefont
  {J.}~\bibnamefont {B{\"{o}}ttcher}}, \bibinfo {author} {\bibfnamefont
  {G.}~\bibnamefont {Tkachov}}, \bibinfo {author} {\bibfnamefont {E.~M.}\
  \bibnamefont {Hankiewicz}}, \bibinfo {author} {\bibfnamefont
  {C.}~\bibnamefont {Br{\"{u}}ne}}, \bibinfo {author} {\bibfnamefont
  {H.}~\bibnamefont {Buhmann}},\ and\ \bibinfo {author} {\bibfnamefont {L.~W.}\
  \bibnamefont {Molenkamp}},\ }\bibfield  {title} {\bibinfo {title}
  {{Observation of the universal magnetoelectric effect in a 3D topological
  insulator}},\ }\href {https://doi.org/10.1038/ncomms15197} {\bibfield
  {journal} {\bibinfo  {journal} {Nat. Commun.}\ }\textbf {\bibinfo {volume}
  {8}},\ \bibinfo {pages} {15197} (\bibinfo {year}
  {2017}{\natexlab{b}})}\BibitemShut {NoStop}%
\bibitem [{\citenamefont {Dziom}(2018)}]{Dziom2018}%
  \BibitemOpen
  \bibfield  {author} {\bibinfo {author} {\bibfnamefont {V.}~\bibnamefont
  {Dziom}},\ }\bibfield  {title} {\bibinfo {title} {{THz spectroscopy of novel
  spin and quantum Hall systems}},\ }\href
  {https://www.ifp.tuwien.ac.at/fileadmin/Arbeitsgruppen/solid_state_spectroscopy/media/thesis_uladzislau_dziom.pdf}
  {\bibfield  {journal} {\bibinfo  {journal} {Ph.D thesis, Vienna Univ.
  Technol.}\ } (\bibinfo {year} {2018})}\BibitemShut {NoStop}%
\bibitem [{\citenamefont {Zhang}\ \emph {et~al.}(2014)\citenamefont {Zhang},
  \citenamefont {Arikawa}, \citenamefont {Kato}, \citenamefont {Reno},
  \citenamefont {Pan}, \citenamefont {Watson}, \citenamefont {Manfra},
  \citenamefont {Zudov}, \citenamefont {Tokman}, \citenamefont {Erukhimova},
  \citenamefont {Belyanin},\ and\ \citenamefont {Kono}}]{Zhang2014b}%
  \BibitemOpen
  \bibfield  {author} {\bibinfo {author} {\bibfnamefont {Q.}~\bibnamefont
  {Zhang}}, \bibinfo {author} {\bibfnamefont {T.}~\bibnamefont {Arikawa}},
  \bibinfo {author} {\bibfnamefont {E.}~\bibnamefont {Kato}}, \bibinfo {author}
  {\bibfnamefont {J.~L.}\ \bibnamefont {Reno}}, \bibinfo {author}
  {\bibfnamefont {W.}~\bibnamefont {Pan}}, \bibinfo {author} {\bibfnamefont
  {J.~D.}\ \bibnamefont {Watson}}, \bibinfo {author} {\bibfnamefont {M.~J.}\
  \bibnamefont {Manfra}}, \bibinfo {author} {\bibfnamefont {M.~A.}\
  \bibnamefont {Zudov}}, \bibinfo {author} {\bibfnamefont {M.}~\bibnamefont
  {Tokman}}, \bibinfo {author} {\bibfnamefont {M.}~\bibnamefont {Erukhimova}},
  \bibinfo {author} {\bibfnamefont {A.}~\bibnamefont {Belyanin}},\ and\
  \bibinfo {author} {\bibfnamefont {J.}~\bibnamefont {Kono}},\ }\bibfield
  {title} {\bibinfo {title} {{Superradiant Decay of Cyclotron Resonance of
  Two-Dimensional Electron Gases}},\ }\href
  {https://doi.org/10.1103/PhysRevLett.113.047601} {\bibfield  {journal}
  {\bibinfo  {journal} {Phys. Rev. Lett.}\ }\textbf {\bibinfo {volume} {113}},\
  \bibinfo {pages} {047601} (\bibinfo {year} {2014})}\BibitemShut {NoStop}%
\end{thebibliography}%

\pagebreak
\widetext

\newpage
\begin{center}
	\textbf{SUPPLEMENTAL MATERIAL}
\end{center}

\setcounter{figure}{0}
\setcounter{equation}{0}
\renewcommand{\thesection}{S\arabic{section}}
\renewcommand{\theequation} {S\arabic{equation}}
\renewcommand{\thefigure} {S\arabic{figure}}
\renewcommand{\thetable} {S\arabic{table}}

\section{Optical SdH oscillations at other frequencies}
\label{SM2}
\noindent

\begin{figure*}[h]
	\includegraphics[width=1\columnwidth]{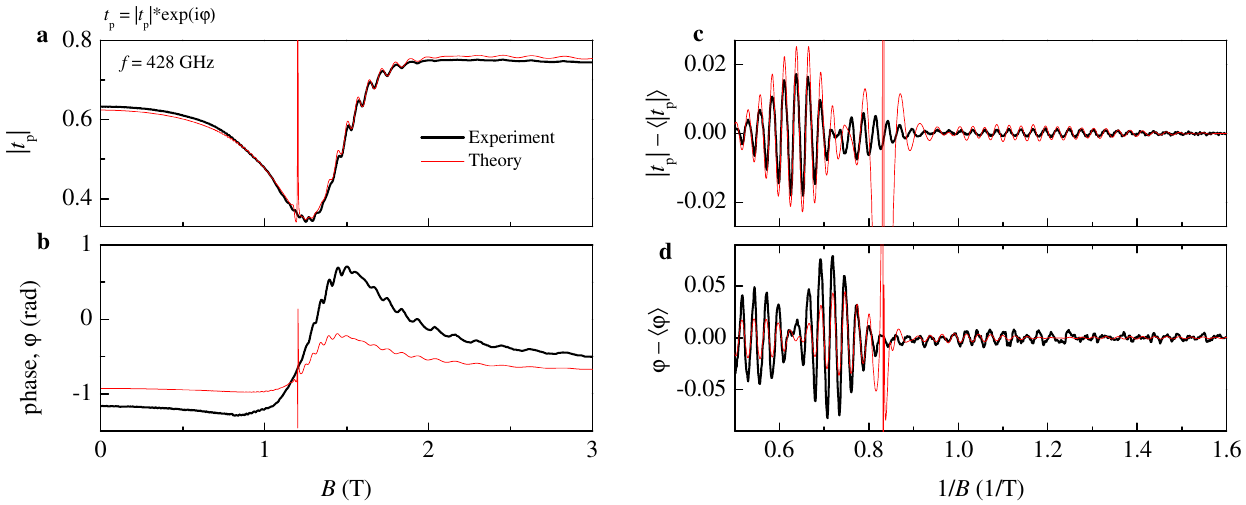}
	\caption{
		(a) and (b) Magnetic field dependencies of the amplitude of the parallel transmittance $|t_\text{p}|$ and phase shift $\varphi$, respectively, measured on GaAs device at 428\,GHz.
		(c)and (d) Oscillations of the parallel transmittance amplitude and phase shift in $1/B$ scale, the smooth parts $\langle |t_\text{p}|^2 \rangle$ and $\langle \varphi \rangle$ are subtracted. 
		Black thick lines present the experimental data, red thin lines are fits based on Eqs.~(1) and (3) of the main text. The series expansion in $\delta$ leading to Eq.~(3) is not justified in the vicinity of the CR, $\mu|B_\text{CR}\pm B|\lesssim \delta$, producing an artificial divergence in theoretical fits around $B = 1.2$~T ($1/B = 0.82$~T$^{-1}$).
	} \label{a020}
\end{figure*}

\begin{figure*}		
	\includegraphics[width=1\columnwidth]{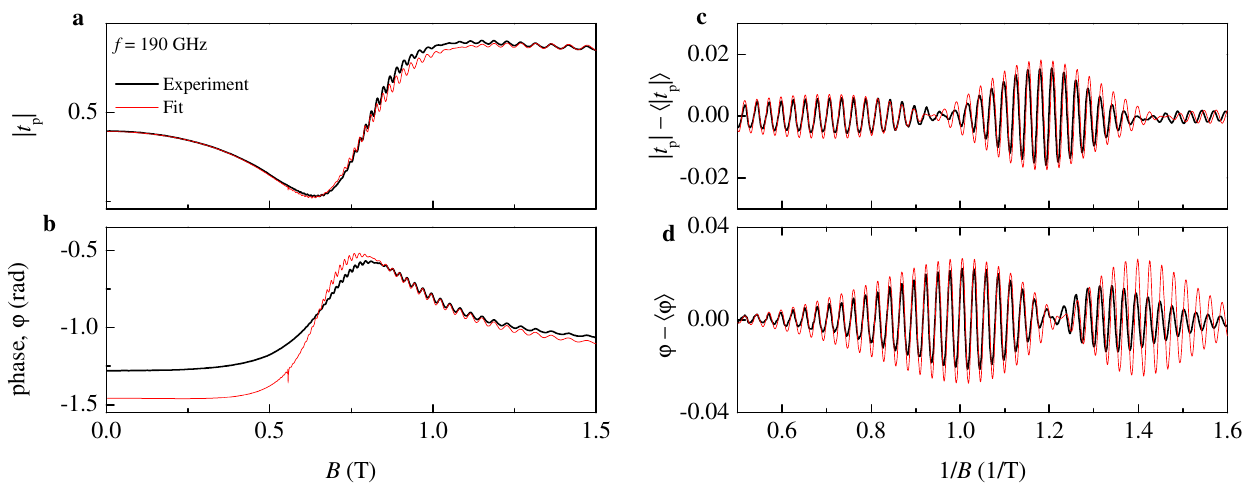}
	\caption{
		(a) and (b) Magnetic field dependences of the amplitude of the parallel transmittance $|t_\text{p}|$ and its phase $\varphi$, respectively, measured on GaAs device at 190\,GHz.
		(c) and (d) The oscillations of the parallel transmittance amplitude and its phase in $1/B$ scale, the smooth parts $\langle |t_\text{p}|^2 \rangle$ and $\langle \varphi \rangle$ are subtracted. 
		Black thick lines present the experimental data, red thin lines are fits based on Eqs.~(1) and (3) of the main text. The series expansion in $\delta$ leading to Eq.~(3) is not justified in the vicinity of the CR, $\mu|B_\text{CR}\pm B|\lesssim \delta$, producing an artificial divergence in theoretical fits around $B = 0.6$\,T.
	} \label{fig4}
\end{figure*}

\newpage
\section{Importance of the imaginary part in the conductivity correction}
\label{SM_Re_test}
\noindent


\begin{figure*}[h]
	\includegraphics[width=0.9\columnwidth]{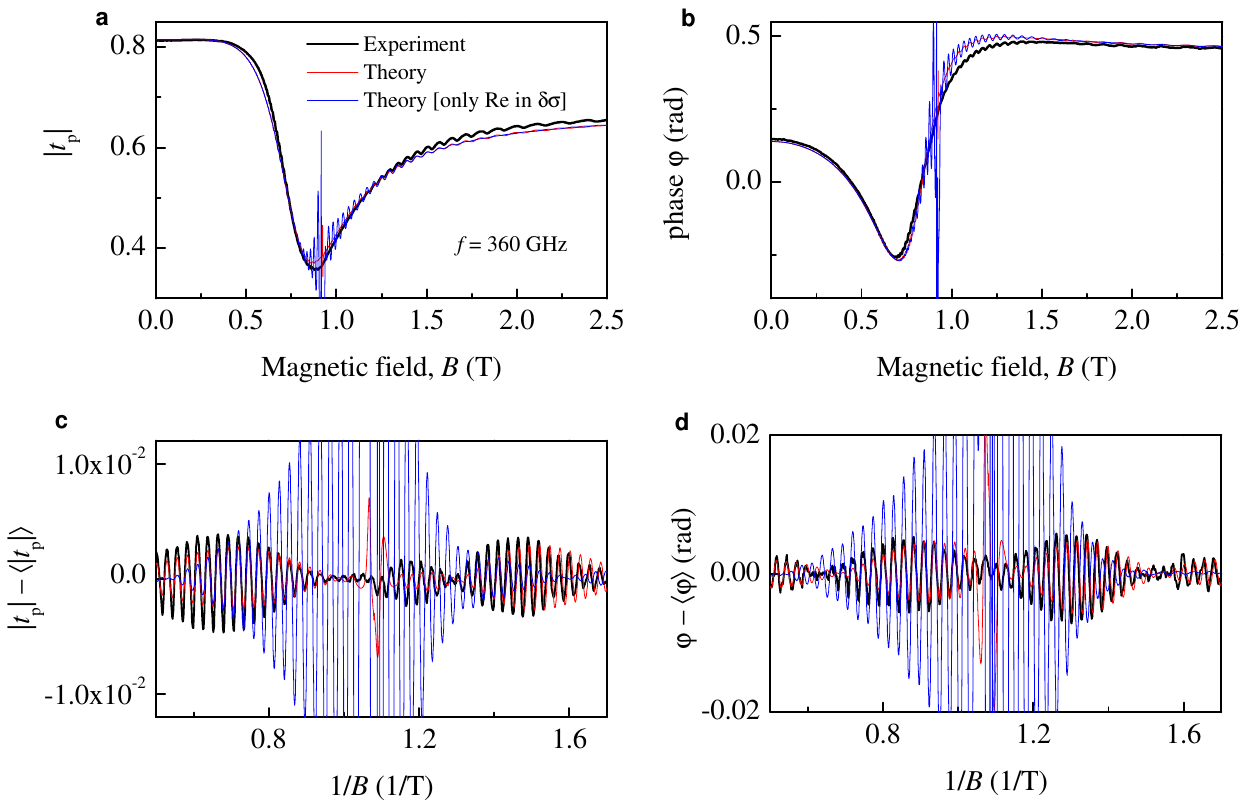}
	\caption{
		Here we replot Fig.~2 of the main text with additional blue curves that were calculated using the same parameters but taking only the real part of the conductivity correction $\delta \sigma$ in Eq.~(3) into account. It is seen that the form of oscillations and positions of the nodes become essentially different, and no longer follow the experimental curves. This illustrates the importance of both real and imaginary parts of of quantum corrections to dynamic conductivity in treating the transmittance experiments.
	} \label{FFT}
\end{figure*}

\newpage
\section{Standing waves in transmission}
\label{SM_Spectrum}
\noindent

\begin{figure*}[h]
	\includegraphics[width=1\columnwidth]{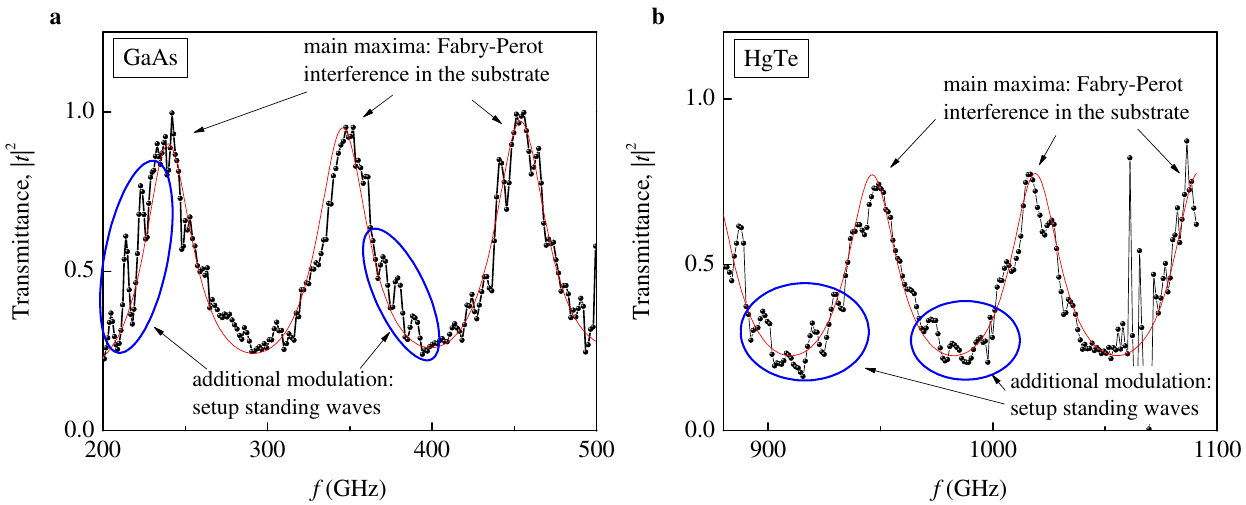}
	\caption{
		Frequency dependence of transmittance $|t|^2$ for the GaAs (a) and HgTe (b) samples measured in the absence of magnetic field. Strong oscillations in both traces come from the Fabry-P\'{e}rot interference in the sample. An additional quasi-periodic modulation originates from standing waves in the optical setup that may modify both frequency and magnetic field dependencies of the transmittance~\cite{Dziom2018}.	The maximum values of the transmittance in the HgTe sample are lower due to the semitransparent metallic gate.} \label{Spectrum}
\end{figure*}

There are two types of standing waves in the transmission signal. The first type is due to Fabry-P\'{e}rot interference in the substrate. It modifies the transmittance according to Eq.~(1) of the main text. The second type is die to inevitable reflections on the optical elements in the setup including cryostat windows, attenuators, lenses, etc. Figure~\ref{Spectrum} shows a typical frequency dependence of the transmittance observed in our experiments. Here the result of both types of the standing waves is seen.
The main maxima come from the interference in the substrate, and they can be fitted by Eq.~(1) (red lines).
An additional modulation, coming from reflections in the optical setup, forms weaker quasi-periodic oscillations.
Apart from distorting the frequency scans, such standing waves modify the magnetic scans as well. 
Therefore, the sample parameters, obtained from a single measurement, may differ from their real values~\cite{Dziom2018}.
To get more precise values, one needs to repeat magnetic field scans at different frequencies and attempt to fit them all with the same set of parameters. 

\newpage
\section{Universal and tunable nodes in transmittance SdH oscillations}
\label{nodes}
\noindent

Nodes in the SdH oscillations of the transmittance can come from two sources. 
First, the SdH-related correction to the optical conductivity (Eq. 3 of the main text)
\begin{equation}
\notag
\delta\sigma \propto f(\alpha_\pm)\cos\left(\frac{ 2\pi^2\hbar n}{e B}\right) \sin \left( \frac{\pi \omega}{\omega_\text{c}} \right) \exp \left( \frac{ i \pi\omega}{ \omega_\text{c}}\right)
\end{equation}
itself has nodes when $\omega/\omega_c$ is equal to an integer number and corresponding $\sin( \pi \omega/\omega_c)$ is equal to zero.

Additionally, since both the Drude transmittance $t_\text{D} = |t_\text{D}|e^{i\varphi_\text{D}}$ and its oscillatory correction $\delta t = |\delta t|e^{i\varphi}$ are complex numbers, they can be perpendicular to each other on a complex plane that results in additional nodes in the measured transmittance. 
Indeed, having the transmittance oscillations much smaller than the Drude transmittance %
\begin{equation}
\notag
|t|^2 = \left ||t_\text{D}|e^{i\varphi_\text{D}} + |\delta t|e^{i\varphi} \right|^2 \approx |t_\text{D}|^2 + 2|t_\text{D}| |\delta t|\cos(\varphi_\text{D} - \varphi).
\end{equation}
When the phases $\varphi_\text{D}$  and $\varphi$ are shifted by $\pi/2$, the transmittance oscillations acquire additional nodes, which we call tunable. In these conditions, variation of magnetic field rotates the complex transmittance $t$ while its length $|t|$ stays the same. To find positions of all tunable nodes in $|t|$ one needs to find zeros of $\cos(\varphi_\text{D} - \varphi)$. 
Taking into account that $f(\alpha_\pm)$ is approximately real away from $B=B_\text{CR}$, the equation determining positions of these nodes is
\begin{equation}
\label{eq: nodeExpressSM}
\tan \left( \frac{\pi\omega}{\omega_c} \right) = \frac{\sqrt{\epsilon} + \tan(\varphi_\text{D})\tan(kd)}{\sqrt{\epsilon}\tan(\varphi_\text{D}) - \tan(kd)}.
\end{equation}
For constructive Fabry-P\'{e}rot interference in the substrate and away from the cyclotron resonance, the right part of this equation becomes a linear function of the magnetic field, $2(B - B_\text{CR})/(n e Z_0)$.\\

%
\begin{figure}[b]		
	\includegraphics[width=1\columnwidth]{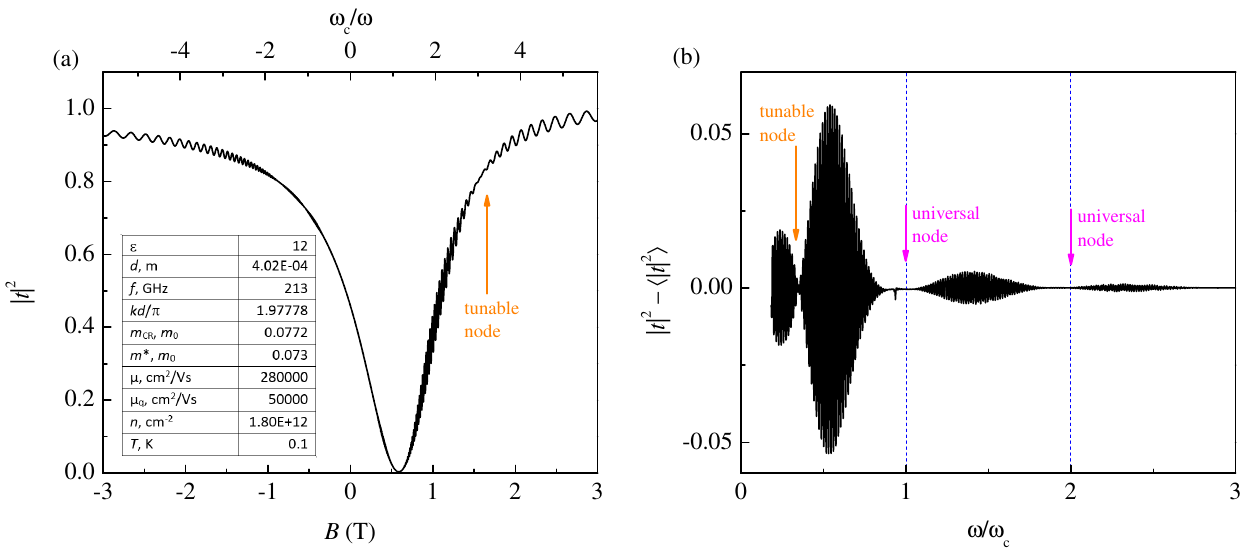}
	\caption{
		(a)~Magnetic field dependence of the transmittance, $|t|^2$ calculated using parameters from the insert.
		(b)~Transmittance oscillations in $\omega/\omega_c$ scale, the smooth part $\langle|t|^2\rangle$ of the transmittance is subtracted.
	} \label{Fig_Nodes1}
\end{figure}
To illustrate the importance of both universal and tunable nodes, 
in Fig.~\ref{Fig_Nodes1}~(a) we show the calculated $|t|^2(B)$ using parameters in the insert. The chosen parameters are close to the ones of studied GaAs system, but to see more oscillations, the temperature is set close to zero and the quantum mobility is taken to be closer to the transport mobility.
Here we have the CR near $B = 0.58$~T and strong optical SdH oscillations.
In panel (b), we show the oscillatory part of the transmittance, $|t|^2 - \langle|t|^2\rangle$ versus $\omega/\omega_c$.
It is clearly seen that there are few universal nodes at $B \le B_{CR}$ and at least one tunable node at $B \approx 1.7~$T.

%
\begin{figure}
	\includegraphics[width=1\columnwidth]{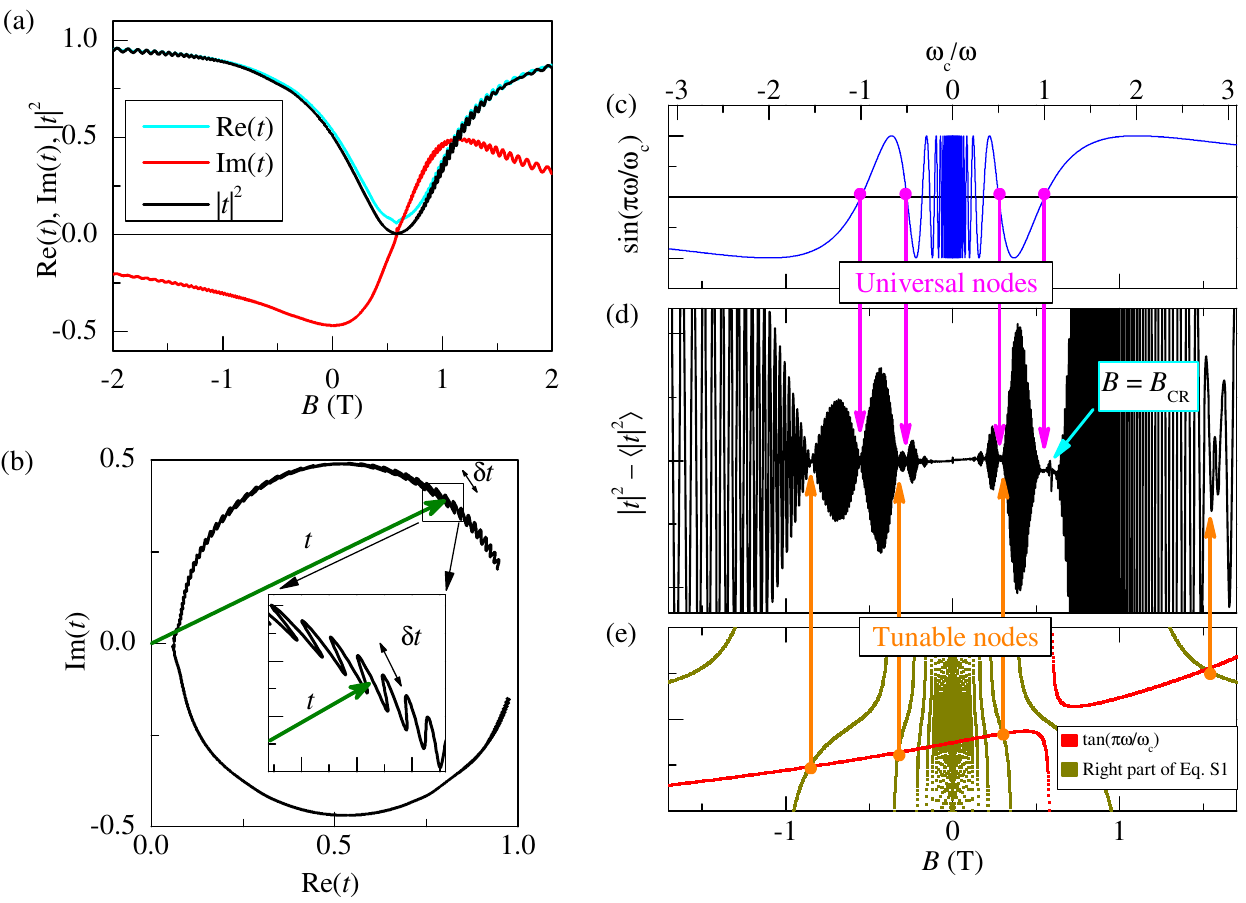}
	\caption{
		(a)~Magnetic field dependences of $|t|^2$ and of the real and imaginary parts of $t$.
		(b)~Imaginary vs real part of the transmission amplitude $t$. 
		The tunable nodes arise when $t$ and its correction $\delta t$ are perpendicular to each other.
		(c), (d) and (e) Magnetic field dependences of $\sin(\pi \omega / \omega_c)$ (origin of the universal nodes), the transmittance SdH oscillations, and graphical solution of Eq.~\eqref{eq: nodeExpressSM} describing the tunable nodes in the transmittance.
	} \label{Fig_Nodes3}
\end{figure}
In Fig.~\ref{Fig_Nodes3}~(a) we show the corresponding real and imaginary parts of $t$ and $|t|^2$.
It is seen that the tunable node in $|t|^2$ at $B = 1.7$ does not correspond to vanishing Re($t$) or Im$(t)$.
In the panel (b) we show the imaginary part of $t$ versus its real part.
The dependence has a usual circular-like Drude behaviour with the CR near minimal values of Re$(t)$.
The optical SdH are also seen, and the total vector $t$ is shown by a green arrow for the magnetic field $B = 1.7~$T equal to the position of the tunable node mentioned above.
It is seen here that at this magnetic field two complex numbers $t$ and $\delta t$ are perpendicular to each other resulting in the formation of the tunable node in $|t|^2$.

In Fig.~\ref{Fig_Nodes3}~(d) we show zoomed transmittance oscillations.
They exhibit universal nodes coming from $\sin(\pi \omega/\omega_c)$~(see panel~(c)) and tunable nodes arising at roots of Eq.~\eqref{eq: nodeExpressSM}~(see panel~(e)).
It is seen that apart from the tunable node at $B = 1.7~$T that we saw before, there are plenty of others.
%
%

Another peculiarity of the optical transmittance oscillations is related to the conditions for their observation. Here it is essential that the optical response is sensitive to conductivity corrections only not too far from the CR ~\cite{Abstreiter1976, Savchenko2020b}.  Away from the CR, when $|\sigma Z_0| \ll 1$ in Eq.~(1), the transmittance becomes insensitive to the conductivity of 2DES. This limits the range of the magnetic fields where optical SdH oscillations can be seen, and this is the main reason to use GaAs quantum well of high density -- the CR width is proportional to the density for high-mobility 2DES~\textcolor{blue}{\cite{Zhang2014b}}.


\end{document}